\documentclass[twocolumn, twocolappendix, trackchanges]{aastex631}
\usepackage{amsmath}

\received{September 28, 2021}
\revised{January 25, 2022}
\accepted{January 31, 2022}

\shorttitle{AMBER}
\shortauthors{Trac et al.}

\begin{document}

\title{AMBER: A Semi-numerical Abundance Matching Box for the Epoch of Reionization}

\correspondingauthor{Hy Trac}
\email{hytrac@andrew.cmu.edu}

\author{Hy Trac}
\affiliation{McWilliams Center for Cosmology, Department of Physics, Carnegie Mellon University, Pittsburgh, PA 15213, USA}
\affiliation{NSF AI Planning Institute for Physics of the Future, Carnegie Mellon University, Pittsburgh, PA 15213, USA}

\author{Nianyi Chen}
\affiliation{McWilliams Center for Cosmology, Department of Physics, Carnegie Mellon University, Pittsburgh, PA 15213, USA}

\author{Ian Holst}
\affiliation{Department of Astronomy and Astrophysics, University of Chicago, Chicago, IL 60637, USA}

\author{Marcelo A.~Alvarez}
\affiliation{Berkeley Center for Cosmological Physics, Department of Physics, University of California, Berkeley, CA 94720, USA}
\affiliation{Lawrence Berkeley National Laboratory, One Cyclotron Road, Berkeley, CA 94720, USA}

\author{Renyue Cen}
\affiliation{Department of Astrophysical Sciences, Princeton University, Princeton, NJ 08544, USA}

\begin{abstract}
The Abundance Matching Box for the Epoch of Reionization (AMBER) is a semi-numerical code for modeling the cosmic dawn. The new algorithm is not based on the excursion set formalism for reionization, but takes the novel approach of calculating the reionization-redshift field $z_\mathrm{re}(\boldsymbol{x})$ assuming that hydrogen gas encountering higher radiation intensity are photoionized earlier. Redshift values are assigned while matching the abundance of ionized mass according to a given mass-weighted ionization fraction $\bar{x}_\mathrm{i}(z)$. The code has the unique advantage of allowing users to directly specify the reionization history through the redshift midpoint $z_\mathrm{mid}$, duration $\Delta_\mathrm{z}$, and asymmetry $A_\mathrm{z}$ input parameters. The reionization process is further controlled through the minimum halo mass $M_\mathrm{min}$ for galaxy formation and the radiation mean free path $l_\mathrm{mfp}$ for radiative transfer. We implement improved methods for constructing density, velocity, halo, and radiation fields, which are essential components for modeling reionization observables. We compare AMBER with two other semi-numerical methods and find that our code more accurately reproduces the results from radiation-hydrodynamic simulations. The parallelized code is over four orders of magnitude faster than radiative transfer simulations and will efficiently enable large-volume models, full-sky mock observations, and parameter-space studies. AMBER will be made publicly available to facilitate and transform studies of the EoR.
\end{abstract}


\section{Introduction} \label{sec:introduction}

Cosmic dawn is a fascinating period in the first billion years of the Universe and is a frontier topic for both theoretical and observational explorations and discoveries. The reionization of hydrogen by the first stars, galaxies, and quasars drastically converts the cold and neutral gas into a warm and highly ionized medium. Galaxies most likely provided the bulk of the ionizing photons \citep[e.g.][]{2015ApJ...811..140B, 2015ApJ...810...71F}, but there may have been early contributions from Population III stars \citep[e.g.][]{2021arXiv210508737W} and late contributions from active galactic nuclei \citep[AGN; e.g.][]{2015ApJ...813L...8M}. On large scales, the higher-density regions near radiation sources are generally reionized earlier than the lower-density regions far from sources in this inhomogeneous process \citep[e.g.][]{2004ApJ...609..474B, 2008ApJ...689L..81T}.

Our current understanding of when the Epoch of Reionization (EoR) occurred primarily comes from ``A Tale of Two Optical Depths". The Thomson optical depth $\tau$ (CMB photons scattering with free electrons) and the Gunn-Peterson optical depth $\tau_\text{GP}$ (Lyman-alpha photons scattering with neutral hydrogen) have long provided two major observational constraints. However, the latest measurements tell a starkly different tale than the early observations \citep[e.g.][]{2002AJ....123.1247F, 2003ApJS..148..161K}.
\citet{2020A&A...641A...6P} recently inferred $\tau =0.054\pm 0.007$ from measurements of the CMB temperature and polarization angular power spectra, implying a late reionization midpoint at redshift $z \approx 7.7 \pm 0.6$ \citep[e.g.][]{2018RNAAS...2c.135G}. There are now multiple evidence of dark Ly$\alpha$ troughs extending down to $z \approx 5.5$ in the spectra of high-redshift quasars \citep[e.g.][]{2015MNRAS.447.3402B, 2018MNRAS.479.1055B, 2018ApJ...864...53E, 2021arXiv210803699B}. This suggests that reionization could have ended at $z < 6$ \citep[e.g.][]{2020MNRAS.491.1736K, 2020MNRAS.tmp..155N}, later than previously assumed. When and how the EoR occurred exactly (i.e.~the reionization history and process) are of primary interest.

There is a renewed emphasis on using radiation-hydrodynamic simulations \citep[e.g.][]{2014ApJ...793...29G, 2015ApJS..216...16N, 2016MNRAS.463.1462O, 2017MNRAS.472.4508S,  2018MNRAS.480.2628F, 2019ApJ...870...18D, 2020ApJ...898..149D, 2021MNRAS.507.1254K} to model the complex astrophysics of sources and sinks on small scales and using semi-analytical/numerical methods \citep[e.g.][]{2004ApJ...613....1F, 2007ApJ...669..663M, 2007ApJ...654...12Z, 2009ApJ...703L.167A, 2009MNRAS.394..960C, 2010MNRAS.406.2421S, 2011MNRAS.411..955M, 2013ApJ...776...81B, 2016MNRAS.457.1550H,  2017ApJ...844..117X, 2018MNRAS.477.1549H} to make theoretical predictions and mock observations on large scales. Radiative transfer (RT) simulations are usually limited to small box sizes $L \lesssim 100\ h^{-1}\mathrm{Mpc}$ and a single one typically requires hundreds of thousands to millions of CPU hours to run on a supercomputer \citep[e.g.][]{2011ASL.....4..228T}. Fast and accurate approaches with larger box sizes are essential for parameter-space studies. These two approaches in tandem provide our best option in making forward progress in synergy with observations. 

A variety of independent approaches are crucial for comparison with and interpretation of observations of the 21cm signal, CMB, Ly$\alpha$ forest, and intensity mapping. However, there has been a lack of development and diversity in semi-analytical/numerical methods. The majority of codes, like the often-used 21cmFAST \citep{2011MNRAS.411..955M}, are based on the excursion set formalism \citep[ESF; e.g.][]{1991ApJ...379..440B, 2004ApJ...613....1F}. In \citet{2011MNRAS.414..727Z}, we compare two RT simulations \citep{2007MNRAS.377.1043M, 2007ApJ...671....1T} with two ESF methods \citep{2007ApJ...669..663M, 2011MNRAS.411..955M}. While the simulations statistically agree within 10\% at all redshifts, the models agree within 50\% only when compared at the same ionization fraction. There are larger statistical differences when compared at the same redshifts because of the disagreement in reionization histories. Previous semi-numerical methods have input astrophysical parameters such as the star formation efficiency, photon production efficiency, and radiation escape fraction. The output reionization history is an end product after many complex and uncertain calculations, and is not easily controlled. There should be flexibility to directly choose the reionization history as this is the primary question for studies of the EoR.

In this paper, we present a new semi-numerical Abundance Matching Box for the Epoch of Reionization (AMBER) code for modeling the cosmic dawn on large scales. AMBER allows users to directly specify the reionization history through input parameters, a useful feature for theoretical and inference studies that is not currently in other semi-numerical methods. Section \ref{sec:scorch} summarizes the Simulations and Constructions of the Reionization of Cosmic Hydrogen (SCORCH) project that is used to motivate and calibrate AMBER. Section \ref{sec:amber} describes the semi-numerical methods in AMBER, including the novel technique for matching the abundance of ionized mass or volume. Section \ref{sec:comparison} compares AMBER against two other semi-numerical methods, and Section \ref{sec:results} presents some basic results. Further applications will be presented in upcoming work. For example, \citet{2022arXiv220304337C} models and studies the patchy kinetic Sunyaev-Zel'dovich (KSZ) effect, which is a promising probe of the EoR. We conclude in Section \ref{sec:conclusion} and add supplementary material in the Appendix.
\section{SCORCH} \label{sec:scorch}

\begin{deluxetable*}{lCCCCCCCCCCCCCC}[t] \label{tab:scorch}
\tablewidth{\textwidth}
\tablecaption{RadHydro and AMBER Parameters}
\tablehead{
\colhead{Model} & \colhead{$L\ [h^{-1}\mathrm{Mpc}]$} & \colhead{$N_\mathrm{dm}$} & \colhead{$N_\mathrm{gas}$} & \colhead{$N_\mathrm{rt}$} &
\colhead{$f_8$} & \colhead{$a_8$} & \colhead{$\tau$} & \colhead{$z_\mathrm{mid}$} &  \colhead{$\Delta_\mathrm{z}$} &  \colhead{$A_\mathrm{z}$} &
\colhead{$a_\mathrm{w}$} &
\colhead{$b_\mathrm{w}$} &
\colhead{$c_\mathrm{w}$} &
\colhead{$l_\mathrm{mfp}$}
}
\startdata
Sim 0 & 50 & $2048^3$ & $2048^3$ & $512^3$ & 0.15 & 0 & 0.060 & 7.95 & 4.68 & 2.90 & 6.62 & 1.85 & 1.14 & 3.0 \\
Sim 1 & & & & & 0.13 & 1 & 0.060 & 7.91 & 5.45 & 2.69 & 6.24 & 2.26 & 1.20 & 3.0 \\
Sim 2 & & & & & 0.11 & 2 & 0.060 & 7.83 & 6.54 & 2.33 & 5.54 & 3.01 & 1.33 & 3.0
\enddata
\end{deluxetable*}

We first summarize the SCORCH project \citep[][]{2015ApJ...813...54T, 2019ApJ...870...18D, 2020ApJ...905..132C} that is used to motivate and calibrate AMBER. SCORCH is designed to provide cosmological simulations, theoretical predictions, and mock observations to facilitate more accurate comparisons with current and future observations. In this section, we describe the N-body simulations (Sec.~\ref{sec:nbody_scorch}), the radiation-hydrodynamic simulations (Sec.~\ref{sec:radhydro_scorch}), and the reionization-redshift fields (Sec.~\ref{sec:zre_scorch}). The simulations are based on the concordance cosmological parameters: $\Omega_\mathrm{b} = 0.045$, $\Omega_\mathrm{m} = 0.3$, $\Omega_\Lambda = 0.7$, $h=0.7$, $\sigma_8 = 0.8$, and $n_\mathrm{s} = 0.96$.

\subsection{N-body Simulations}
\label{sec:nbody_scorch}

In SCORCH I \citep{2015ApJ...813...54T}, we  run 22 high-resolution N-body simulations to quantify the abundance of dark matter halos as a function mass $M$, accretion rate $\dot{M}$, and redshift $z$ during the EoR. The N-body simulations are run using an updated particle-particle-mesh (P$^3$M) code with a hybrid halo finder. Box sizes in the range $10 \leq L/(h^{-1}\mathrm{Mpc}) \leq 400$ are chosen to focus on the atomic cooling halos and two realizations of each box size are run to reduce sample variance. Each simulation contains $N_\mathrm{dm} = 2048^3$ dark matter particles and has a particle mass resolution $m_\mathrm{p} = 8.72 \times 10^6(L/100)^3\ h^{-1}M_\odot$.

We quantify and fit the halo mass function $dn/dM$, mass accretion-rate relation $\dot{M}(M,z)$, and halo accretion-rate function $dn/d\dot{M}$ for the redshift range $z \ge 6$. The new fit for the halo mass function is $20-40\%$ more accurate compared to \citet{2008ApJ...688..709T} at the high-mass end hosting currently observable galaxies. We also model and study the galaxy-halo connection by abundance matching observed high-redshift galaxies with simulated dark matter halos. See \citet{2015ApJ...813...54T} for more details.

In SCORCH II \citep{2019ApJ...870...18D}, we also run a high-resolution P$^3$M simulation with $N_\text{dm} = 3072^3$ dark matter particles in a comoving box of side length $L = 50\ h^{-1}\text{Mpc}$ to generate halo and galaxy catalogs for the radiation-hydrodynamic simulations. A halo finder is run on the fly every 20 million cosmic years to locate dark matter halos and build merger trees. With a particle mass resolution of $m_\text{p} = 3.59 \times 10^5\ h^{-1}M_\odot$, we can reliably measure halo quantities such as mass and accretion rate down to the atomic cooling limit for galaxy formation ($T \sim 10^4$ K, $M \sim 10^8\ h^{-1}M_\odot$).

\subsection{Radiation-hydrodynamic Simulations} \label{sec:radhydro_scorch}

In SCORCH II \citep{2019ApJ...870...18D}, we run three radiation-hydrodynamic simulations with the same cosmic initial conditions, same galaxy luminosity functions, but with different radiation escape fraction $f_{\rm esc}(z)$ models. The simulations are designed to have the same Thomson optical depth $\tau \approx 0.06$, consistent with recent CMB observations \citep{2020A&A...641A...6P}, and similar midpoints of reionization $7.5 \lesssim z \lesssim 8$, but with different evolution of the ionization fraction $\bar{x}_\text{i}(z)$.

The radiation-hydrodynamic simulations are run with the RadHydro code, which combines N-body and hydrodynamic algorithms \citep{2004NewA....9..443T} with an adaptive raytracing algorithm \citep{2007ApJ...671....1T} to directly and simultaneously solve collisionless dark matter dynamics, collisional gas dynamics, and RT of ionizing photons. The raytracing algorithm has adaptive splitting and merging to improve resolution and scaling. Each simulation has $N_\text{dm} = 2048^3$ dark matter particles, $N_\text{gas} = 2048^3$ gas cells, $N_\mathrm{rt} = 512^3$ RT cells, and up to 12 billion adaptive rays in a comoving box of side length $50\ h^{-1}\text{Mpc}$. RadHydro has been used to simulate both hydrogen and helium reionization \citep[e.g.][]{2008ApJ...689L..81T, 2013ApJ...776...81B, 2017ApJ...841...87L, 2020ApJ...898..149D}.

We model high-redshift galaxies using an updated approach that allows us to systematically control the galaxy distributions in the simulations while matching the observed luminosity functions from HST \citep[e.g.][]{2015ApJ...803...34B, 2015ApJ...810...71F}. We populate dark matter halos with galaxies by abundance matching the number densities,
\begin{equation}
n_{\rm galaxy}(> L_{\rm UV},z) = n_{\rm halo}(>\dot{M},z) ,
\end{equation}
where $L_{\rm UV}$ is the galaxy UV luminosity and $\dot{M}$ is the halo mass accretion rate. Connecting the mass accretion rate to the star formation rate allows us to model and study the episodic nature of high-redshift galaxy formation.

The radiation escape fraction is allowed to vary with redshift in a power-law relation,
\begin{equation}
f_{\rm esc}(z) = f_8 \left(\frac{1+z}{9}\right)^{a_8} ,
\end{equation}
where $f_8$ is the average escape fraction at $z=8$, and $a_8$ is the power-law slope. Sim ($a_8=$) 0 has constant $f_{\rm esc}$ and reionization starts latest but ends earliest out of the three models. Sim 1 has $f_{\rm esc}(z)$ varying linearly with $1+z$ and is an intermediate model. Sim 2 has $f_{\rm esc}(z)$ varying quadratically and reionization starts earliest but ends latest.

Table \ref{tab:scorch} summarizes the parameters for the three RadHydro simulations. Some additional parameters will be presented later in the paper: the reionization history parameters $z_\mathrm{mid}$, $\Delta_\mathrm{z}$, and $A_\mathrm{z}$ in Sec.~\ref{sec:zDA_amber}, the Weibull coefficients $a_\mathrm{w}$, $b_\mathrm{w}$, and $c_\mathrm{w}$ in Sec.~\ref{sec:Weibull_amber}, and the radiation mean free path $l_\mathrm{mfp}$ in Sec.~\ref{sec:radfield_amber}. Each simulation takes approximately half a million CPU hours to run on a supercomputer, and they are infeasible for parameter-space studies. See SCORCH I and II \citep{2015ApJ...813...54T, 2019ApJ...870...18D} for more details.

\subsection{Reionization-redshift Fields} \label{sec:zre_scorch}

\begin{figure*}[t]
\includegraphics[width=\textwidth]{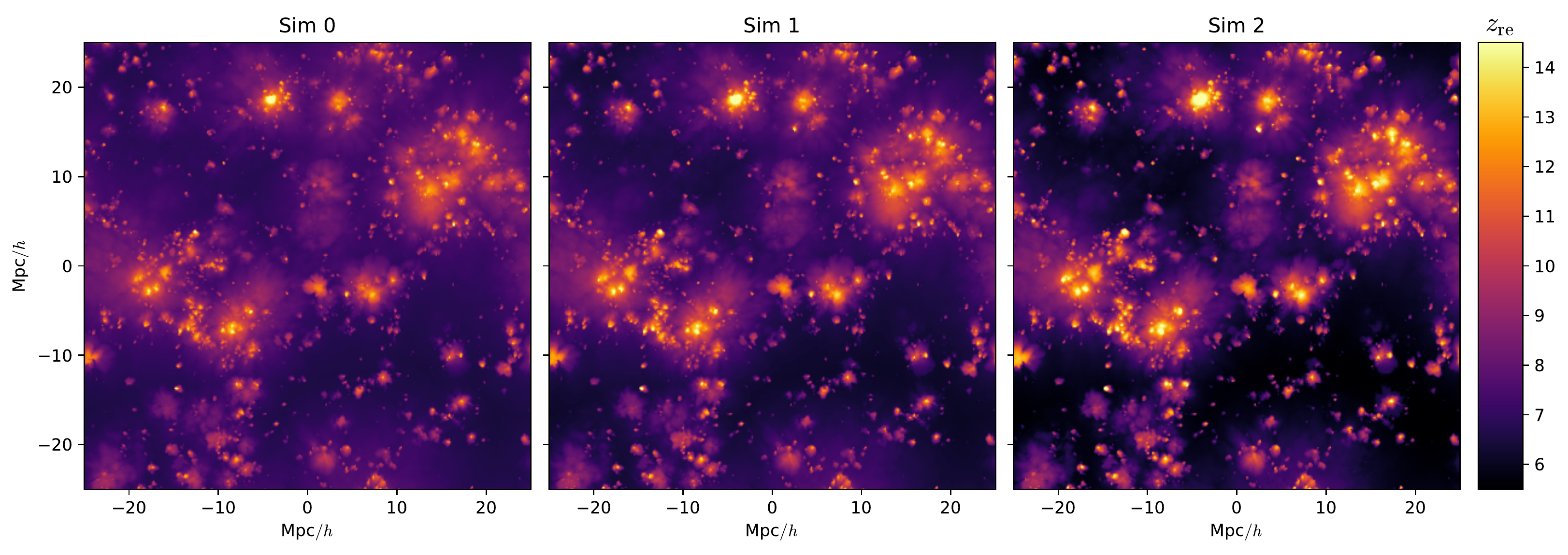}
\caption{Visualization of the reionization-redshift fields $z_\mathrm{re}(\boldsymbol{x})$ from three RadHydro simulations \citep[SCORCH II;][]{2019ApJ...870...18D}. Each image is $512 \times 512$ pixels from a slice that is $50\, h^{-1}\mathrm{Mpc} \times 50\, h^{-1}\mathrm{Mpc}$ with a thickness of $\sim 100\, h^{-1}\mathrm{kpc}$.  The three simulations show remarkable similarities even with episodic star formation and varying radiation escape fractions. On large scales, higher-density regions near sources are generally reionized earlier than lower-density regions far from sources.}
\label{fig:zre_scorch}
\end{figure*}

The reionization-redshift field $z_\mathrm{re}(\boldsymbol{x})$ quantifies the timing of reionization as a function of space and has been measured in radiation-hydrodynamic simulations \citep[e.g.][]{2008ApJ...689L..81T, 2013ApJ...776...81B, 2016ApJ...831..198K, 2019ApJ...870...18D} and semi-analytical/numerical models \citep[e.g.][]{2004ApJ...609..474B, 2009ApJ...703L.167A, 2014MNRAS.443.2843M}. Combined with the density, velocity, and temperature fields, it has been used to model the imprints of reionization on the CMB \citep[e.g.][]{2013ApJ...776...82N, 2013ApJ...776...83B}, 21cm \citep[e.g.][]{2014ApJ...789...31L, 2019ApJ...880..110L}, and the Ly$\alpha$ forest \citep[e.g.][]{2015ApJ...813L..38D, 2019ApJ...874..154D}.

We construct the reionization-redshift field for a given RadHydro simulation while it is running. For each gas cell, we save the redshift at which the majority of the mass is ionized in a three-dimensional array. Previously in \citet{2013ApJ...776...81B}, we studied both 50\% and 90\% ionization thresholds and find no significant differences at the resolutions of interest. Statistically, the autocorrelations and cross correlations of the $z_\mathrm{re}(\boldsymbol{x})$ fields for these two cases are almost identical. For the vast majority of gas cells, once they start to ionize they quickly become almost fully ionized, but not entirely because of recombinations. Thus, we adopt a nominal 50\% ionization value for the RadHydro simulations in SCORCH II \citep{2019ApJ...870...18D} and throughout this paper.

Figure \ref{fig:zre_scorch} is a visualization of the reionization-redshift fields $z_\mathrm{re}(\boldsymbol{x})$ from the three RadHydro simulations. They show remarkable similarities even with episodic star formation $\dot{\rho}_\star(\boldsymbol{x},z)$ and varying radiation escape fractions $f_\mathrm{esc}(z)$. The $z_\mathrm{re}(\boldsymbol{x})$ field is highly correlated with the matter density field $\rho(\boldsymbol{x})$ on large scales since the higher-density regions near sources are generally reionized earlier than the lower-density regions far from sources \citep[e.g.][]{2004ApJ...609..474B, 2008ApJ...689L..81T}. In \citet{2013ApJ...776...83B}, we find that these two fields are correlated down to Mpc scales and can be related using a first-order, scale-dependent bias in Fourier space.

\subsubsection{Cross Correlations}

\begin{figure}[t]
\includegraphics[width=\linewidth]{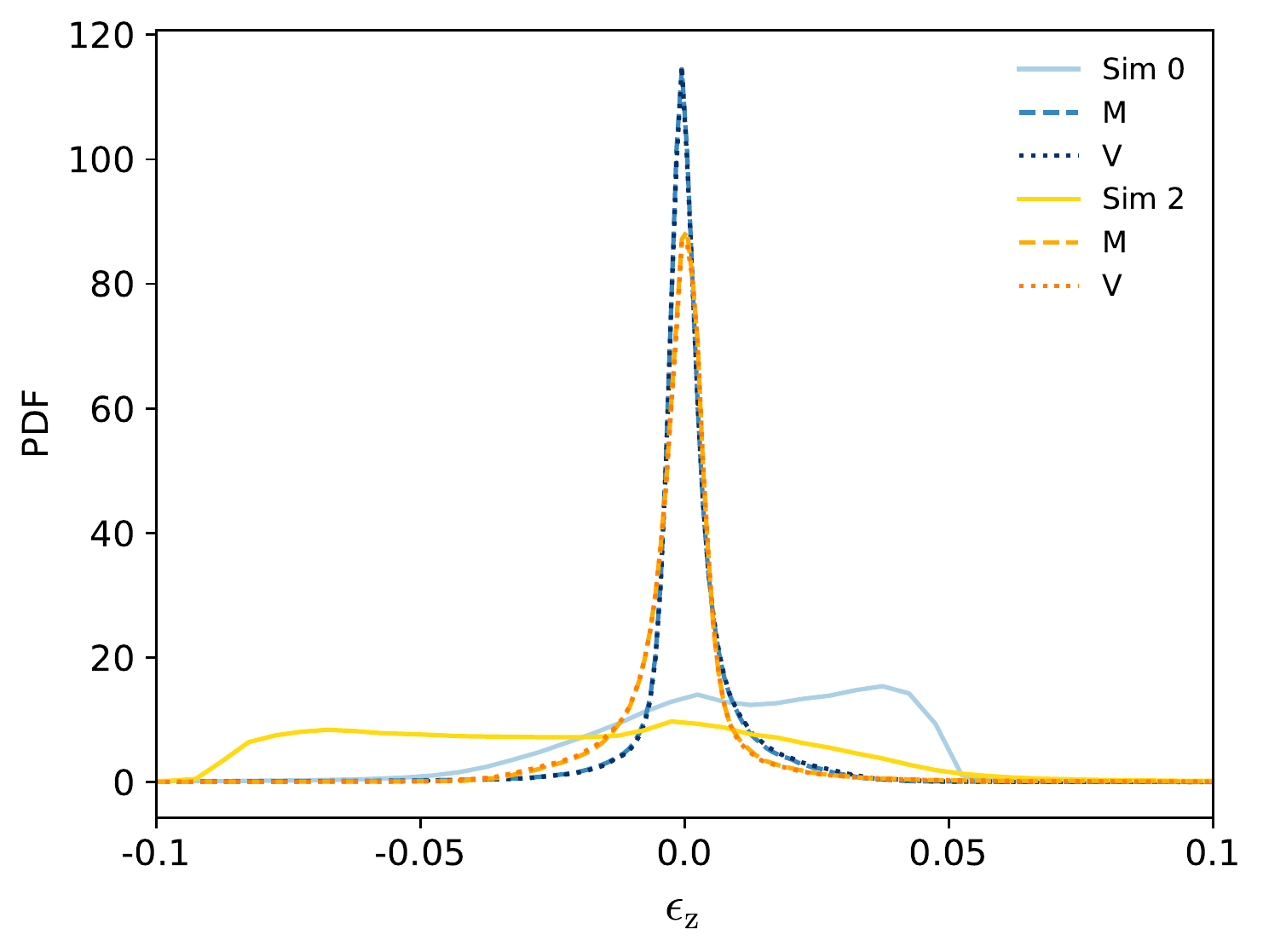}
\caption{Distribution of fractional differences $\epsilon_\mathrm{z}$  in the reionization-redshift fields $z_\mathrm{re}(\boldsymbol{x})$ for the RadHydro Sims 0 and 2 relative to Sim 1. Prior to abundance matching (solid), the distributions are skewed because of the different reionization histories. After abundance matching to have the same mass-weighted (M) or volume-weighted (V) ionization fractions, the distributions are approximately Gaussian with zero mean and small widths.}
\label{fig:zre_err_scorch}
\end{figure}

\begin{figure}[t]
\includegraphics[width=\linewidth]{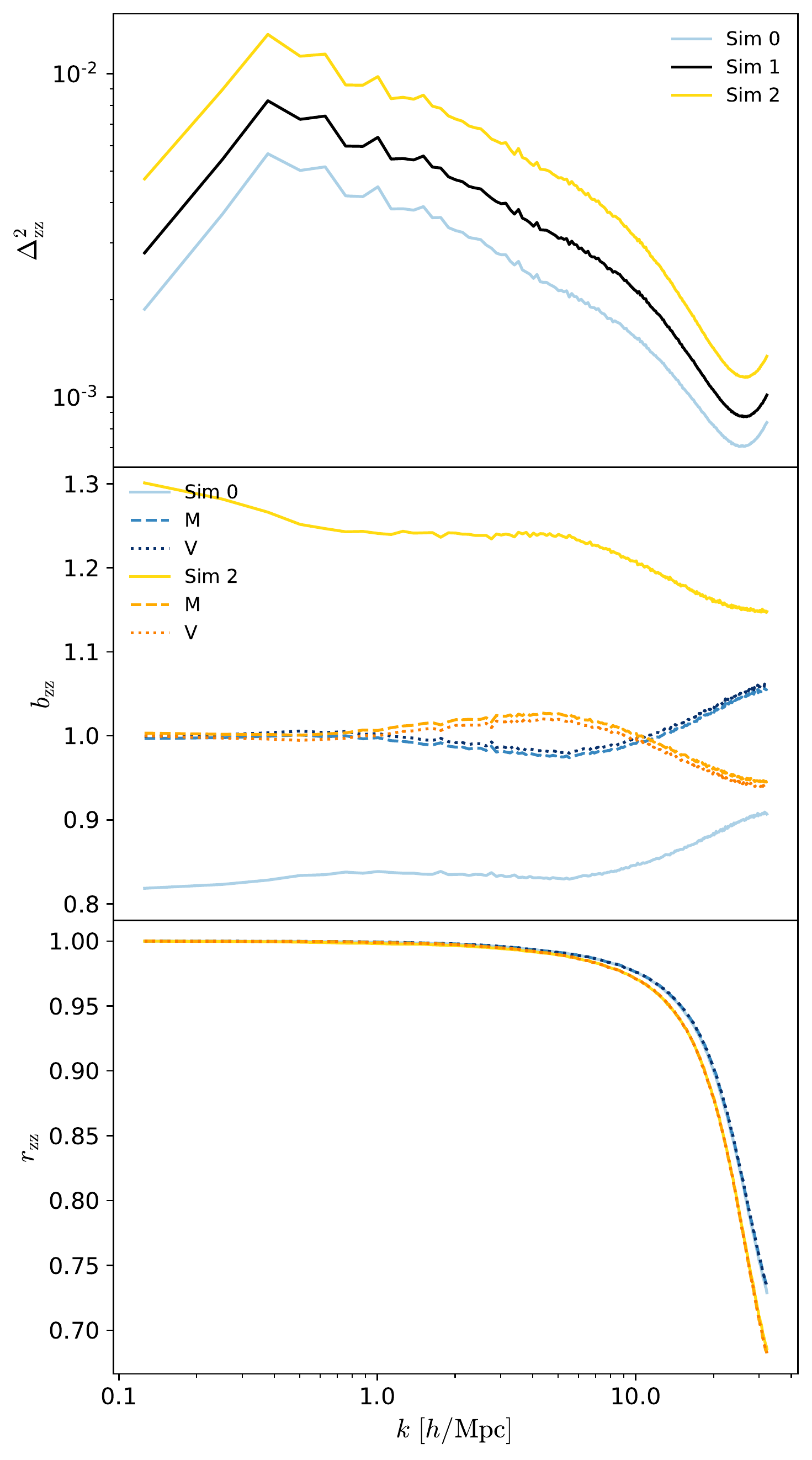}
\caption{{\bf Top:} autopower spectra of the RadHydro reionization-redshift fields $z_\mathrm{re}(\boldsymbol{x})$ prior to abundance matching. All of the spectra have a characteristic peak on larger scales and a power-law break on smaller scales. {\bf Center:} the redshift bias $b_\mathrm{zz}$ is nonunity when crosscorrelating the original $z_\mathrm{re}(\boldsymbol{x})$ fields, but is much closer to unity after abundance matching to have the same mass-weighted (M) or volume-weighted (V) ionization fractions. {\bf Bottom:} the cross correlation $r_\mathrm{zz}$ shows that the original fields are already very highly correlated on scales $k \lesssim 10\ h/\mathrm{Mpc}$ and abundance matching still preserves this correlation.}
\label{fig:zre_power_scorch}
\end{figure}

The reionization-redshift fields are quantitatively very similar when we account for the different reionization histories. To show this, we take the $z_\mathrm{re}(\boldsymbol{x})$ fields from Sims 0 and 2 and individually scale the redshift values for each grid cell without changing their spatial rank order such that they have the same ionization fraction $\bar{x}_{0,2}(z) = \bar{x}_{1}(z)$ as Sim 1. We are matching the abundance of ionized mass or volume at all redshifts, which we call abundance matching the reionization history (Sec.~\ref{sec:zream_amber} for more details). \citet{2012ApJ...747..126A} use a similar technique to rescale the reionization-redshift field in their semi-numerical method. Here we consider both mass-weighted $\bar{x}_\mathrm{i,M}(z)$ and volume-weighted $\bar{x}_\mathrm{i,V}(z)$ ionization fractions.

To compare the $z_\mathrm{re}(\boldsymbol{x})$ fields cell by cell, we define the fractional difference field,
\begin{equation} \label{eqn:epsz}
    \epsilon_\mathrm{z}(\boldsymbol{x}) \equiv \frac{[1+z_\mathrm{re}(\boldsymbol{x})] - [1+z_1(\boldsymbol{x})]}{1+z_1(\boldsymbol{x})} ,
\end{equation}
for Sim 0 or 2 relative to Sim 1 and then calculate the probability distribution function (PDF) of $\epsilon_\mathrm{z}$. Figure \ref{fig:zre_err_scorch} shows the distribution of fractional differences for the $z_\mathrm{re}(\boldsymbol{x})$ fields at the RT resolution $l_\mathrm{rt} = 98\ h^{-1}\mathrm{kpc}$, which is approximately ten times smaller than the typical mesh cell size used in semi-numerical models. The difference distributions for the original Sims 0 and 2 are skewed because the reionization histories differ from that of Sim 1. After abundance matching, the distributions are approximately Gaussian with zero mean and small root-mean-square differences $\sigma_\mathrm{z} = \langle (\epsilon_\mathrm{z}(\boldsymbol{x}) - \bar{\epsilon}_\mathrm{z})^2 \rangle^{1/2} \lesssim 0.01$. We find very similar results when matching either the mass-weighted or the volume-weighted ionization fractions.

To correlate the fields, we define the normalized redshift field,
\begin{equation} \label{eqn:deltaz}
    \delta_\mathrm{z}(\boldsymbol{x}) \equiv \frac{[1+z_\mathrm{re}(\boldsymbol{x})] - [1 + \bar{z}_\mathrm{re}]}{1 + \bar {z}_\mathrm{re}} ,
\end{equation}
where $\bar{z}_\mathrm{re}$ is the volume-weighted average value, and Fourier transform to get $\delta_\mathrm{z}(\boldsymbol{k})$ for all three simulations. Generally, the two-point power spectrum and its dimensionless version are defined as
\begin{align}
    \label{eqn:Pk}
    P_{ij}(k) & \equiv \langle\delta_i(\boldsymbol{k})\delta_j(\boldsymbol{k})\rangle , \\
    \label{eqn:Dk}
    \Delta_{ij}^2(k) & \equiv \frac{k^3}{2\pi^2}P_{ij}(k) ,
\end{align}
where $i=j$ for autocorrelations and $i \ne j$ for cross correlations. For comparing two different fields, we quantify with the bias and cross correlation,
\begin{align}
    \label{eqn:bias}
    b_{ij}(k) & \equiv \sqrt{\frac{P_{ii}(k)}{P_{jj}(k)}} , \\
    \label{eqn:rcc}
    r_{ij}(k) & \equiv \frac{P_{ij}(k)}{\sqrt{P_{ii}(k)P_{jj}(k)}} .
\end{align}
The field $\delta_i$ is said to be biased, unbiased, or underbiased relative to $\delta_j$ for $b_{ij} > 1$, $= 1$, and $< 1$, respectively. Similarly, the fields and their fluctuations are said to be correlated, uncorrelated, and anticorrelated for $r_{ij} > 0$, $= 0$, and $< 0$, with perfect correlation or anticorrelation given by the limits $r_{ij} = 1$ and $= -1$, respectively.

Figure \ref{fig:zre_power_scorch} shows the autopower spectra for the $z_\mathrm{re}(\boldsymbol{x})$ fields prior to abundance matching. On larger scales, the dimensionless spectra similarly peak at $k \approx 0.3-0.4\ h/\text{Mpc}$. As this is only a factor of 3 from the RadHydro simulation box limit $k_\mathrm{min} = 2\pi/L$, the peak scale may possibly be underestimated. Nonetheless, this suggests that reionization simulations and semi-numerical models must have box lengths $> 20\ h^{-1}\mathrm{Mpc}$ in order to capture the relevant large-scale fluctuations and correlations in the reionization-redshift field. The spectra similarly decline in power-law form on intermediate scales until $k \approx 10\ h/\text{Mpc}$, followed by steeper decline on smaller scales down to the RT Nyquist frequency.

We also crosscorrelate Sims 0 and 2 relative to Sim 1 and plot the bias $b_\mathrm{zz}$ and normalized cross correlation $r_\mathrm{zz}$ in Figure \ref{fig:zre_power_scorch}. Using the normalized fields $\delta_\mathrm{z}(\boldsymbol{x})$ partially adjusts for the different redshift midpoints, but the shorter and longer durations result in lower and higher bias for the original Sims 0 and 2, respectively. After abundance matching, the $b_\mathrm{zz}(k)$ curves are unity at the box scale and deviate by only $\lesssim 10\%$ at the RT resolution scale. The cross correlations show that the original fields are already very highly correlated, allowing abundance matching to be applied correctly. The $r_\mathrm{zz}(k)$ curves are exactly unity at the largest scale and remain within a few percent on scales $k \lesssim 10\ h/\mathrm{Mpc}$.

All of these results lead to new ideas for AMBER: there is a spatial order to the reionization process, and abundance matching can be applied to a correlated field to accurately predict the reionization-redshift field.
\section{AMBER} \label{sec:amber}

The AMBER code provides a novel semi-numerical method for modeling the cosmic dawn on large scales. The new algorithm is not based on the ESF for directly predicting the ionization fraction field \citep[e.g.][]{2004ApJ...613....1F}, but takes the novel approach of calculating the reionization-redshift field $z_\mathrm{re}(\boldsymbol{x})$ assuming that the hydrogen gas encountering higher radiation intensity are photoionized earlier. Redshift values are assigned while matching the abundance of ionized mass according to a given mass-weighted ionization fraction $\bar{x}_\mathrm{i}(z)$. The code has the unique advantage of allowing users to directly specify the reionization history through input parameters, a useful feature for theoretical and inference studies. In this section, we describe the major model components: reionization history (Sec.~\ref{sec:reionhistory_amber}), Lagrangian perturbation theory (LPT) for the large-scale structure (Sec.~\ref{sec:lpt_amber}), ESF for the halo mass density field (Sec.~\ref{sec:esf_amber}), and abundance matching for the reionization-redshift field (Sec.~\ref{sec:abundmatch_amber}).

\subsection{Reionization History} \label{sec:reionhistory_amber}

\begin{figure}[t]
\includegraphics[width=\hsize]{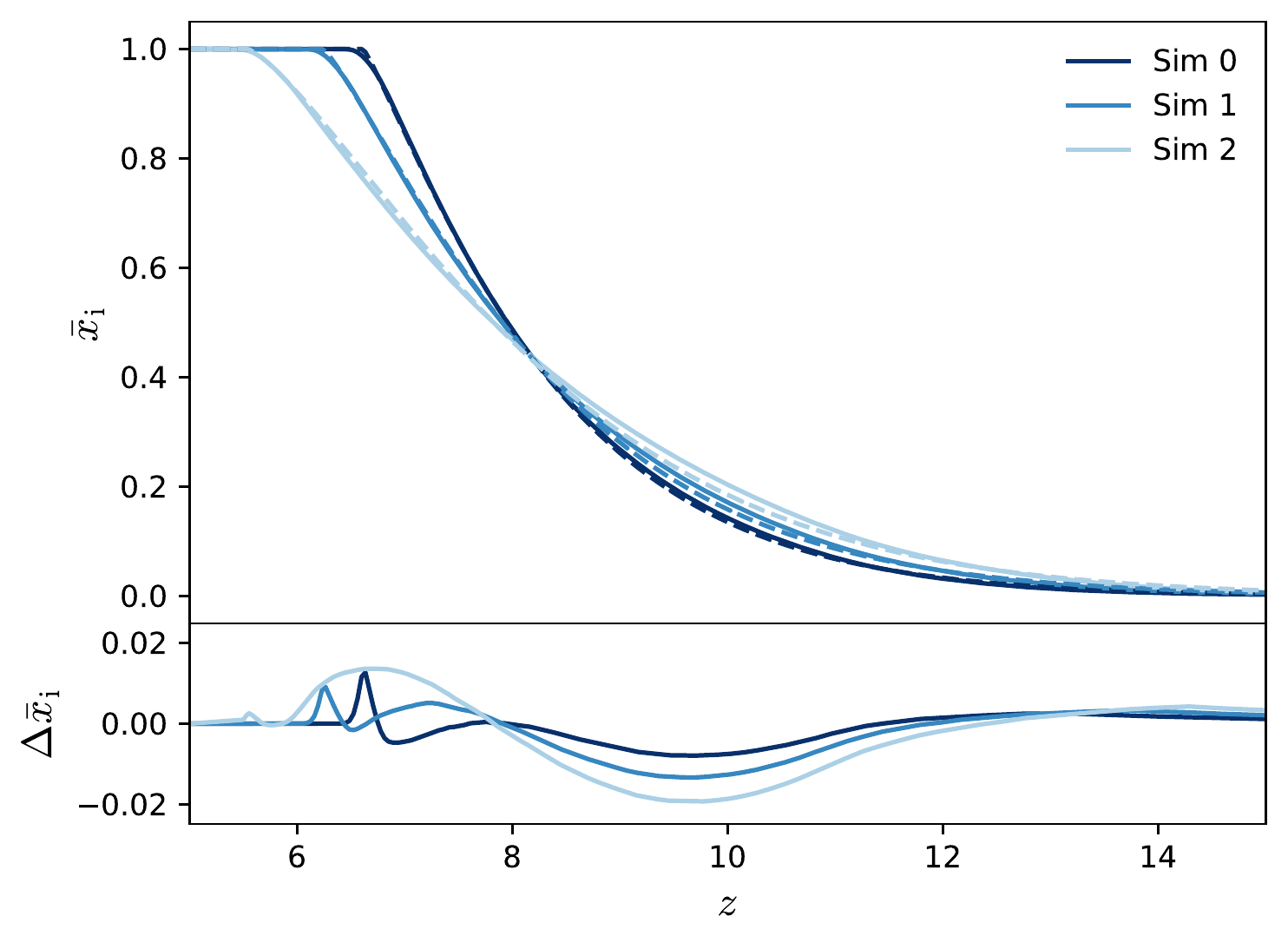}
\caption{{\bf Top:} the evolution of the mass-weighted ionization fraction with redshift from the RadHydro simulations (solid) and the Weibull functions (dashed). The analytical parameterizations accurately capture the redshift-asymmetric form of the simulated reionization histories. {\bf Bottom:} the typical differences in ionization fractions are $|\Delta\bar{x}_\mathrm{i}| \lesssim 0.01$, while the maximum differences of $|\Delta\bar{x}_\mathrm{i}| \lesssim 0.02$ are typically found near the end of the EoR, which are uncertain in the simulations.}
\label{fig:xi_scorch}
\end{figure}

The reionization history $\bar{x}_\mathrm{i}(z)$ directly affects EoR observables. For example, the integrated Thomson optical depth and the evolution of the global 21cm brightness temperature depend linearly on the ionized electron fraction $\bar{x}_\mathrm{e}(z)$ and neutral hydrogen fraction $\bar{x}_\mathrm{HI}(z)$, respectively. In \citet{2018ApJ...858L..11T}, we show that the reionization histories of the RadHydro simulations can be effectively and accurately described by the redshift midpoint, duration, and asymmetry parameters. We will here generalize and improve the approach for AMBER.

The reionization history is quantified by the average ionized hydrogen fraction, which can be mass-weighted or volume-weighted. We work with the mass-weighted version $\bar{x}_\mathrm{i,M}$ as the volume-averaged ionized hydrogen number density is given by
\begin{equation}
\bar{n}_\mathrm{HII,V} = \bar{x}_\mathrm{i,M} \bar{n}_\mathrm{H,V} .
\end{equation}
A common misconception is that the volume-averaged ionized density is instead proportional to the volume-weighted ionization fraction. See \citet{2020ApJ...905..132C} for clarification. We will typically work with mass-weighted fractions and volume-averaged densities and drop the subscripts to simplify the notation.

\subsubsection{Midpoint, Duration, and Asymmetry} \label{sec:zDA_amber}

The redshift midpoint $z_\mathrm{mid}$, duration $\Delta_\mathrm{z}$, and asymmetry $A_\mathrm{z}$ are input parameters, which in turn give three ionization points. For the early, middle, and late stages of reionization, let the redshifts $z_\mathrm{ear} > z_\mathrm{mid} > z_\mathrm{lat}$ correspond to the ionization fractions $\bar{x}_\mathrm{ear} < \bar{x}_\mathrm{mid} < \bar{x}_\mathrm{lat}$. We take $z_\mathrm{mid}$ as the redshift midpoint and define the duration,
\begin{equation} \label{eqn:duration}
\Delta_\mathrm{z} \equiv z_\mathrm{ear} - z_\mathrm{lat} ,
\end{equation}
and asymmetry,
\begin{equation} \label{eqn:asymmetry}
A_\mathrm{z} \equiv \frac{z_\mathrm{ear} - z_\mathrm{mid}}{z_\mathrm{mid} - z_\mathrm{lat}} .
\end{equation}
The other two redshifts are then given by
\begin{align}
    \label{eqn:zear}
    z_\mathrm{ear} & = z_\mathrm{mid} + \frac{\Delta_\mathrm{z}A_\mathrm{z}}{1 + A_\mathrm{z}} , \\
    \label{eqn:zlat}
    z_\mathrm{lat} & = z_\mathrm{mid} - \frac{\Delta_\mathrm{z}A_\mathrm{z}}{1 + A_\mathrm{z}} = z_\mathrm{ear} - \Delta_\mathrm{z} .
\end{align} 
Instantaneous reionization models would correspond to $\Delta_\mathrm{z} = 0$, but reionization is an extended process with finite $\Delta_\mathrm{z} > 0$. Symmetric reionization histories would correspond to $A_\mathrm{z} = 1$, but reionization simulations typically find that the early stage of reionization takes longer than the later stage such that $A_\mathrm{z} > 1$.

Table \ref{tab:scorch} lists the midpoint, duration, and asymmetry parameters for the three RadHydro simulations. In \citet{2019ApJ...870...18D}, we take $\bar{x}_\mathrm{mid} = 0.50$ and present two practical choices for the other ionization points. In the first case, we choose quartile ionization fractions $(\bar{x}_\mathrm{ear},\bar{x}_\mathrm{lat}) = (0.25,0.75)$ and $\Delta_\mathrm{z}$ is analogous to a full width half max. In the second case, we take $(\bar{x}_\mathrm{ear},\bar{x}_\mathrm{lat}) = (0.05,0.95)$ and $\Delta_\mathrm{z}$
more effectively quantifies the whole EoR. We will adopt the latter as the default convention throughout this paper. AMBER users can specify other values, but extreme choices (e.g.~$\bar{x}_\mathrm{ear} \lesssim 0.01$, $\bar{x}_\mathrm{lat} \gtrsim 0.99$) are not recommended because the start and end of the EoR are not well defined and are difficult to determine precisely.

\subsubsection{Analytical Interpolating Function} \label{sec:Weibull_amber}

Once the midpoint, duration, and asymmetry parameters are specified and used to calculate three ionization points, we use an analytical interpolating function to calculate the reionization history. In \citet{2018ApJ...858L..11T}, we use Lagrange interpolating functions to construct an analytical function for the ionization fraction $x_\mathrm{i}(z)$ that exactly fits the ionization points. For three points, the interpolating polynomial is a quadratic, which has the advantage of being invertible but has the disadvantage of being nonmonotonic. A log transformation of the variables improves monotonicity over the relevant redshift range, but it is not an ideal solution.

In AMBER, we interpolate the three ionization points with a modified Weibull function \citep{1951JAM....18..293W},
\begin{equation}
    \label{eqn:weibull}
    \bar{x}_\mathrm{i}(z) = \exp\left[-\max\left(\frac{z-a_\mathrm{w}}{b_\mathrm{w}},0\right)^{c_\mathrm{w}}\right] ,
\end{equation}
where the coefficients $a_\mathrm{w}$, $b_\mathrm{w}$, $c_\mathrm{w}$ are all positive values. Note that $a_\mathrm{w}$ corresponds to $z_\mathrm{end}$ when $x_\mathrm{end} = 1$ and the $\max$ function ensures full ionization at lower redshifts. In Appendix \ref{app:weibull}, we show that the coefficients can be easily determined by first solving a nonlinear equation for $c_\mathrm{w}$ and then substituting its value into algebraic equations for the other two coefficients. We find that solutions exist for the asymmetry range $A_\mathrm{z} \lesssim 15$, which is more than sufficient for parameter space studies. We also experimented with a generalized logistic function \citep{doi:10.1093/jxb/10.2.290}, but find that it is limited to $A_\mathrm{z} \lesssim 5$. The Weibull function can be analytically inverted,
\begin{equation}
    z(\bar{x}_\mathrm{i}) = a_\mathrm{w} + b_\mathrm{w}\left|\ln \bar{x}_\mathrm{i}\right|^{1/c_\mathrm{w}} ,
\end{equation}
which is a useful feature for our abundance matching technique.

Table \ref{tab:scorch} lists the Weibull coefficients for the three RadHydro simulations. The offset $a_\mathrm{w}$ determines when the function reaches unity, and it decreases when reionization ends later. The divisor $b_\mathrm{w}$ controls the stretch in redshift, and it increases with the duration $\Delta_\mathrm{z}$. The power $c_\mathrm{w}$ controls the steepness of the function, and it is inversely related to the asymmetry $A_\mathrm{z}$. In general, changing one of the reionization shape parameters while holding the other two fixed affects all three Weibull coefficients.

Figure \ref{fig:xi_scorch} compares the evolution of the mass-weighted ionization fraction $x_\mathrm{i}(z)$ from the RadHydro simulations with the Weibull functions. The analytical parameterizations are excellent matches to the simulated reionization histories. The typical differences are only $|\Delta\bar{x}_\mathrm{i}| \lesssim 0.01$, while the maximum differences of $|\Delta\bar{x}_\mathrm{i}| \lesssim 0.02$ are found near the end of the EoR, which is not accurately captured in reionization simulations and semi-analytical models. We find that the Weibull function gives slightly better fits than Lagrange interpolation and is valid for a larger range of parameter space.

\subsection{Lagrangian Perturbation Theory} \label{sec:lpt_amber}

LPT is used to generate initial conditions for cosmological simulations and semi-analytical methods. LPT is also used in semi-numerical models of reionization to efficiently model the evolved matter distribution in the moderately nonlinear regime because particles can be simply advanced to any given redshift without having to iteratively perform expensive force calculations and time integration like in N-body and hydro simulations. Furthermore, on moderately nonlinear scales, the dark matter and gas distributions are highly correlated and assumed to exactly trace each other. We will here test these assumptions and techniques for evolving the large-scale structure and constructing the density and velocity fields in AMBER.

In first-order LPT, otherwise known as the Zel'dovich Approximation \citep{1970A&A.....5...84Z}, each particle moves in a straight line with displacement and velocity,
\begin{align}
    \boldsymbol{x} & = \boldsymbol{q} - D(z)\nabla\phi(\boldsymbol{q}) , \\[5pt]
    \boldsymbol{v} & = - D(z)f(z)H(z)\nabla\phi(\boldsymbol{q}) ,
\end{align}
where $\boldsymbol{q}$ is the Lagrangian coordinate, $\phi(\boldsymbol{q})$ is the Lagrangian potential, $D(z)$ is the density linear growth factor, $f(z)$ is the velocity linear growth factor, and $H(z)$ is the Hubble function. In second-order LPT \citep[e.g.][]{1995A&A...296..575B, 1998MNRAS.299.1097S}, each particle moves in a slightly curved trajectory with displacement and velocity,
\begin{align} \label{eqn:xlpt}
    \boldsymbol{x} & = \boldsymbol{q} - D_1\nabla\phi_1 + D_2\nabla\phi_2, \\[5pt]
    \boldsymbol{v} & = -D_1f_1 H\nabla\phi_1 + D_2f_2 H\nabla\phi_2,
\end{align}
where $\phi_i(\boldsymbol{q})$, $D_i(z)$, and $f_i(z)$ are the usual terms calculated at the $i$th-order. The spatial information is contained in the Lagrangian potentials, which only have to be computed once, while the time evolution is governed by the growth factors, which are more straightforward to calculate.

For modeling the EoR, we find that 2LPT is more accurate than 1LPT at length scales $\lesssim 1\ \mathrm{Mpc}$ and redshifts $z \gtrsim 5$. Third-order LPT \citep[e.g.][]{1995A&A...296..575B, 1998MNRAS.299.1097S} and augmented LPT \citep[e.g.][]{2013MNRAS.435L..78K, 2016MNRAS.455L..11N} have been shown to offer only small improvements at lower redshifts far after the end of the EoR. Hence, we will generally use second-order theory and will simply refer to it as LPT throughout the remainder of this paper.

\subsubsection{Density Fields} \label{sec:density_amber}

\begin{figure*}[t]
\center
\includegraphics[width=0.9\textwidth]{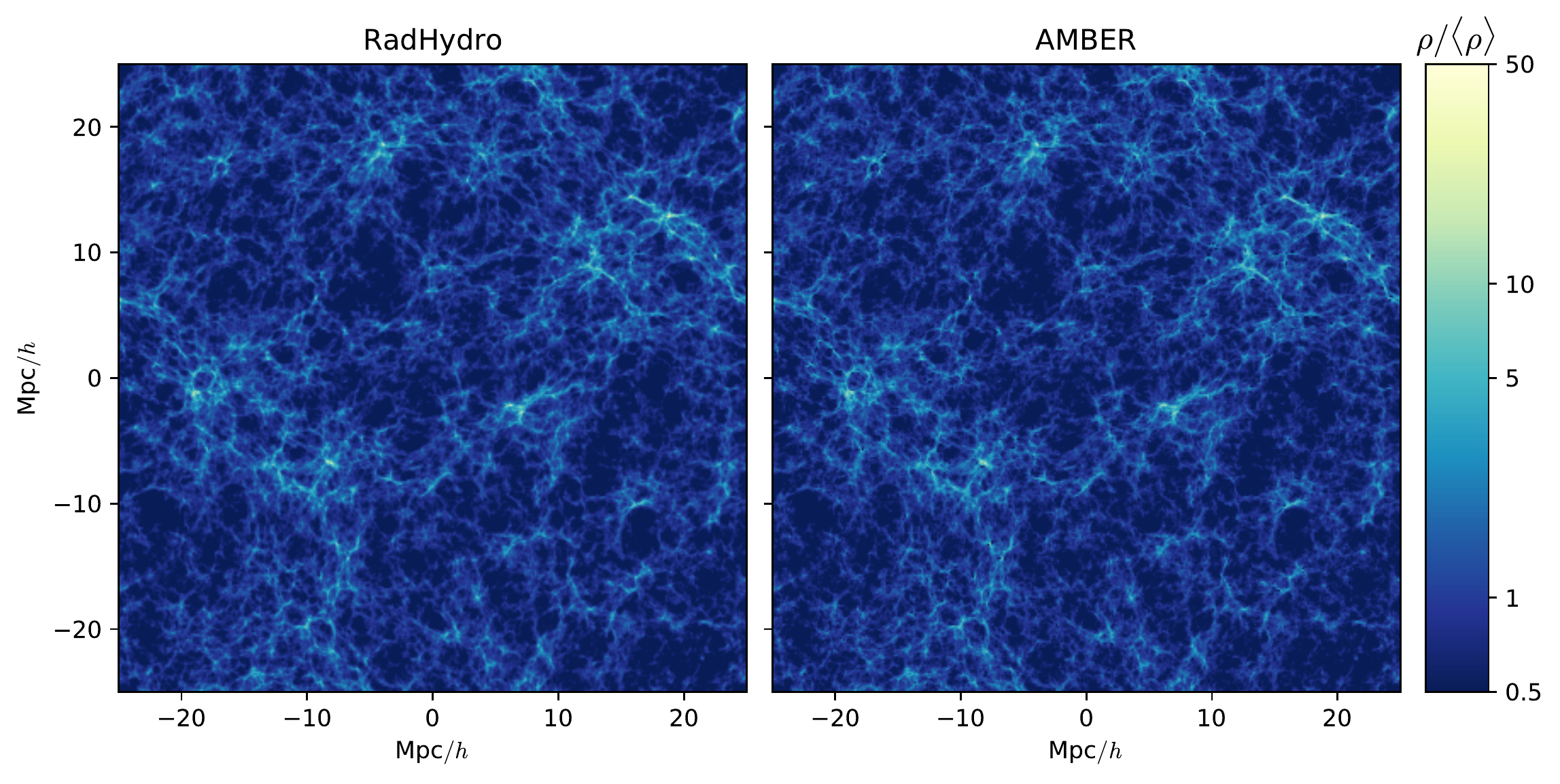}
\caption{Visualization of the matter density fields $\rho(\boldsymbol{x})/\langle\rho\rangle$ at redshift $z=8$ from RadHydro (left) and AMBER (right). Each image is $512 \times 512$ pixels from a slice that is $50\, h^{-1}\mathrm{Mpc} \times 50\, h^{-1}\mathrm{Mpc}$ with a thickness of $\sim 1\, h^{-1}\mathrm{Mpc}$. The RadHydro matter and gas (not shown) distributions are visually indistinguishable on the scales of interest. The AMBER density field, made by assigning the LPT particles using the TSC scheme with interlacing and deconvolution, also closely resembles the RadHydro results.}
\label{fig:rho_image}
\end{figure*}

Density fields are constructed from the LPT particles using techniques from particle-mesh (PM) methods. In N-body simulations, particles are added to a Cartesian mesh to create a density field for solving Poisson's equation with Fast Fourier Transforms (FFT). In large-scale structure analyses, galaxies or halos are added to a grid to create a density field for calculating the power spectrum. Similar techniques are also used in other semi-numerical models of reionization, but we will describe an improved implementation for AMBER.

There is a hierarchy of particle assignment schemes, and the first three are called the nearest grid point (NGP), cloud-in-cell (CIC), and triangular-shaped cells (TSC). The mass assignment process interpolates the discrete distribution of particles and can be interpreted as convolving and sampling the underlying density field $\rho(\boldsymbol{x})$ to produce a discretized density field,
\begin{equation}
    \rho(\boldsymbol{x_\mathrm{n}}) = \int \rho(\boldsymbol{x}) W_\mathrm{pm}(\boldsymbol{x_\mathrm{n}} - \boldsymbol{x}) d^3x,
\end{equation}
where $W_\mathrm{pm}(\boldsymbol{x})$ is the PM assignment window function. See \citet{Hockney:1988:CSU:62815} for a seminal review.

The particle assignments introduce effects such as aliasing, smoothing, and shot noise that have to be accounted for \citep[e.g.][]{2005ApJ...620..559J}. While the latter two effects can have analytical corrections, the first is especially problematic. \citet{Hockney:1988:CSU:62815} have suggested that interlacing two or more staggered meshes can reduce aliasing. Interlacing has been used to improve force calculations in N-body simulations \citep{1991ApJ...368L..23C}, produce more accurate overdensity fields for power spectra estimation \citep{2016MNRAS.460.3624S}, and recently implemented in the nbodykit package \citep{2018AJ....156..160H}.

We use the interlacing technique to produce more accurate and robust density fields in AMBER. We construct the first density field $\rho_1(\boldsymbol{x})$ as normal, while for the second density mesh $\rho_2(\boldsymbol{x})$, the particle positions are shifted by half of a grid cell spacing in each direction by adding the vector $\boldsymbol{\Delta x} = (l_\mathrm{c}/2,l_\mathrm{c}/2,l_\mathrm{c}/2)$, which is equivalent to shifting the mesh by the opposite vector. In Fourier space, the two transformed fields are combined into a single, effective field as
\begin{equation} \label{eqn:interlace}
    \rho(\boldsymbol{k}) = \frac{\left[\rho_1(\boldsymbol{k}) + e^{i\boldsymbol{k}\cdot\boldsymbol{\Delta x}}\rho_2(\boldsymbol{k})\right]/2}{W_\mathrm{pm}(\boldsymbol{k})} .
\end{equation}
The twiddle factor $e^{i\boldsymbol{k}\cdot\boldsymbol{\Delta x}}$ multiplied to the second transformed field accounts for the shifted particle positions. We also deconvolve the effects of smoothing by dividing with the Fourier transform of the assignment window function,
\begin{equation} \label{eqn:Wofk}
    W_\mathrm{pm}(\boldsymbol{k}) = \left[\frac{\sin(k_\mathrm{x}l_\mathrm{c}/2)\sin(k_\mathrm{y}l_\mathrm{c}/2)\sin(k_\mathrm{z}l_\mathrm{c}/2)}{(k_\mathrm{x}l_\mathrm{c}/2)(k_\mathrm{y}l_\mathrm{c}/2)(k_\mathrm{z}l_\mathrm{c}/2)}\right]^p ,
\end{equation}
where $p = 1$, 2, and 3 for the NGP, CIC, and TSC schemes, respectively.

For comparison, we construct the AMBER and RadHydro density fields at two resolutions, using a $64^3$ mesh with cell size $l_\mathrm{c} = 0.8\ h^{-1}\mathrm{Mpc}$ and a $512^3$ mesh with cell size $l_\mathrm{c} = 0.1\ h^{-1}\mathrm{Mpc}$. The fiducial lower resolution is more typical of semi-numerical models, while the higher resolution allows us to test the limits of LPT. The AMBER density fields are efficiently made using equal numbers of LPT particles as mesh cells. The RadHydro density fields are made by simply binning the $2048^3$ dark matter particles and the $2048^3$ gas cells down to the appropriate size meshes.

Figure \ref{fig:rho_image} is a visualization of the matter density fields $\rho(\boldsymbol{x})$ at redshift $z=8$ from RadHydro and AMBER. The LPT particles are assigned using the TSC scheme with interlacing and deconvolution. Both images are shown with the higher-resolution pixels in order to see the large-scale structure more clearly, but have been averaged along the line of sight at the lower resolution. The RadHydro matter and gas (not shown) density fields are visually indistinguishable, while the AMBER density field also has close resemblance, but minor smoothing is visible in the small-scale, high-density regions. At the lower resolution, the AMBER and RadHydro fields are indistinguishable, though the images appear quite pixelated.

\begin{figure}[t!]
\includegraphics[width=\linewidth]{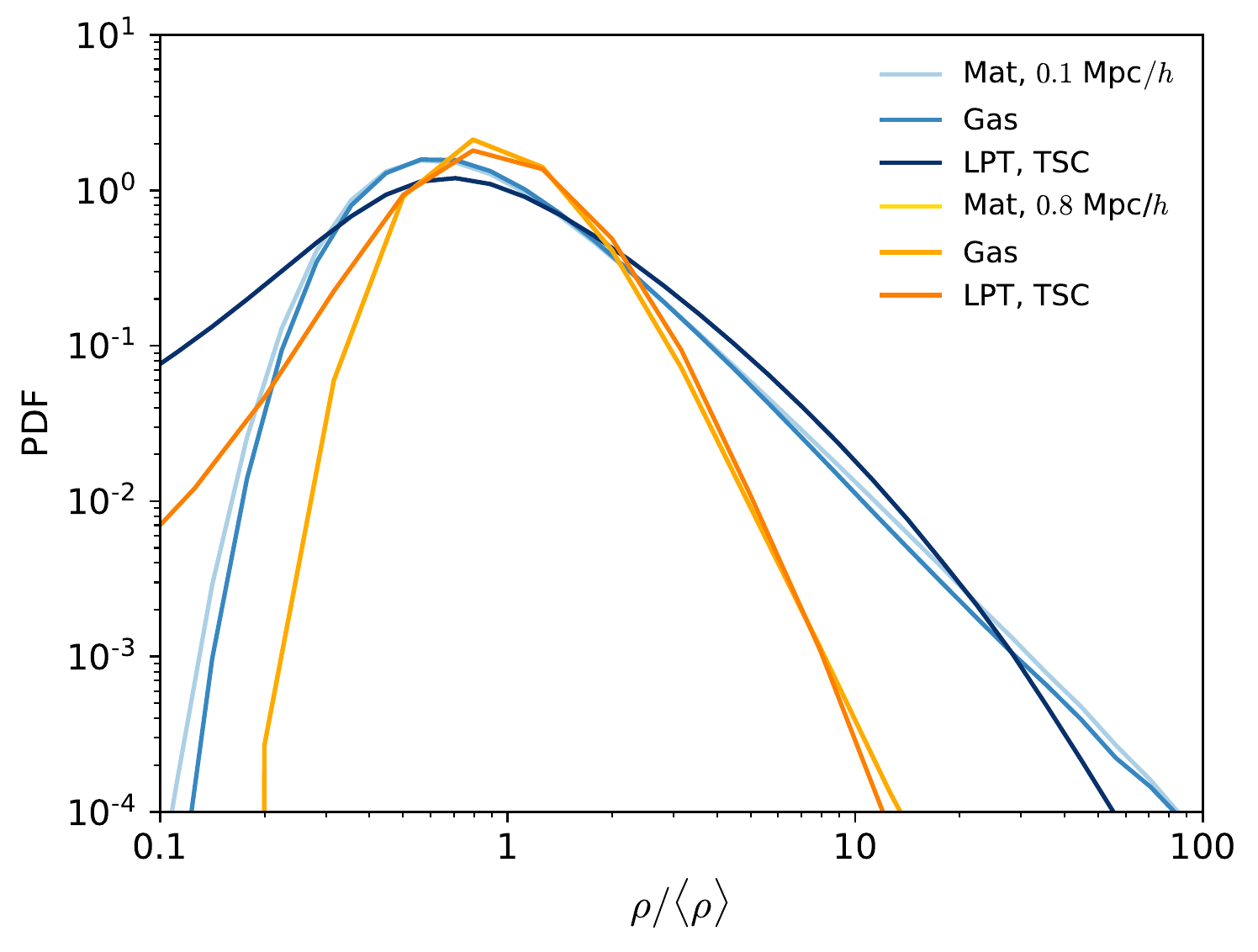}
\caption{One-point probability distribution functions of the density fields at redshift $z=8$ from RadHydro and AMBER. The RadHydro matter and gas distributions are identical at the fiducial lower resolution and in very good agreement at the higher resolution. For AMBER, using LPT with the TSC scheme also gives very good agreement at the fiducial lower resolution. The differences in the high-density tails of the PDFs are due to reduced gravitational collapse and imperfect deconvolution for LPT.}
\label{fig:rho_pdf}
\end{figure}

\begin{figure}[t!]
\includegraphics[width=\linewidth]{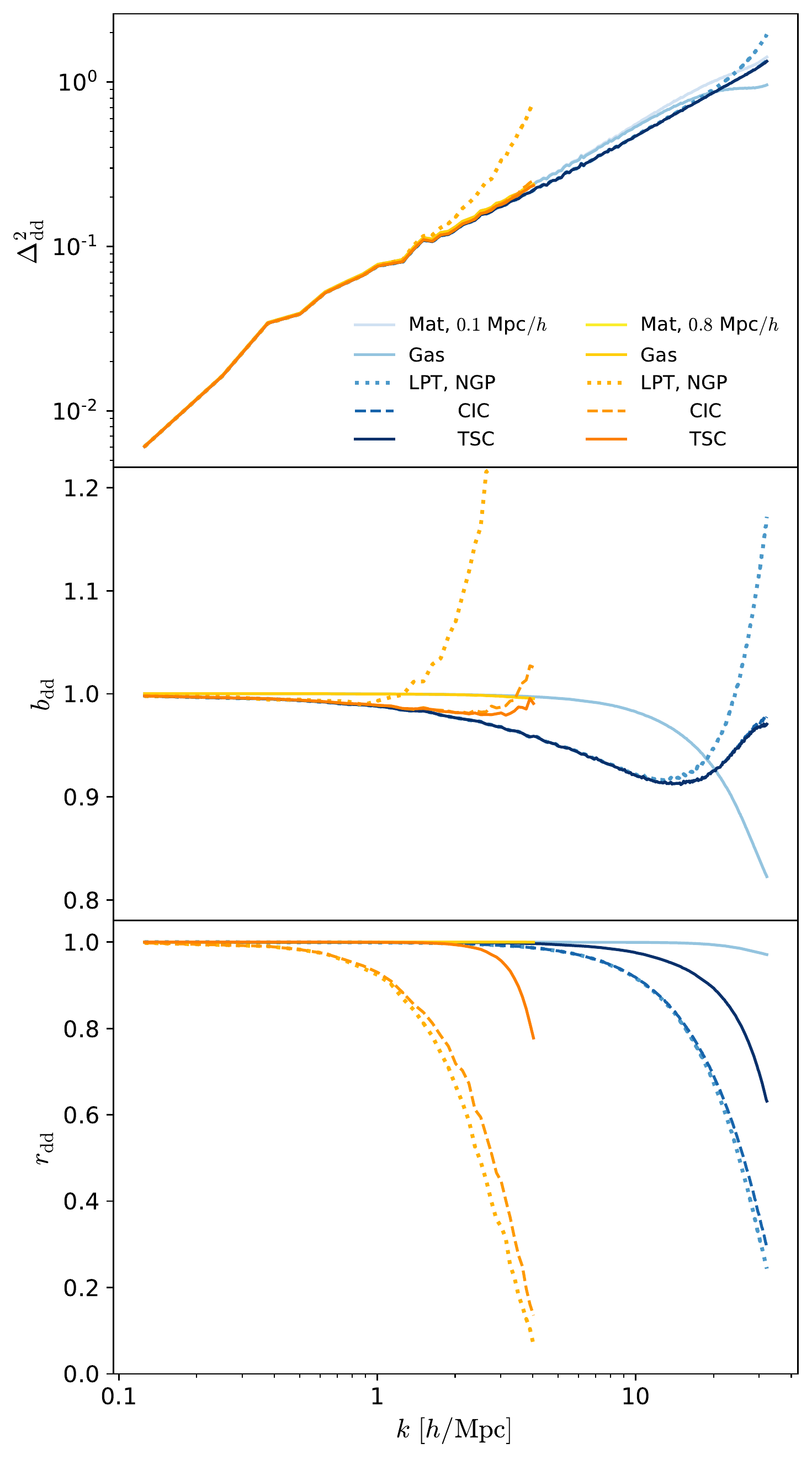}
\caption{{\bf Top:} autopower spectra of the density fields at redshift $z=8$ from RadHydro and AMBER. {\bf Center:} for RadHydro, the density bias $b_\mathrm{dd}$ of the gas relative to the matter is unity for $k \lesssim 10\ h/\mathrm{Mpc}$ and the suppression on smaller scales is due to Jeans smoothing from the gas pressure. For AMBER, the TSC assignment scheme with interlacing and deconvolution gives more correct power than the NGP and CIC schemes, which tend to overshoot near the Nyquist frequency because of aliasing. {\bf Bottom:} the cross correlation $r_\mathrm{dd}$ of the gas relative to the matter is unity nearly down to the smallest scale probed. Assigning LPT particles with the TSC scheme significantly improves the cross correlations.}
\label{fig:rho_cc}
\end{figure}

Figure \ref{fig:rho_pdf} compares the one-point PDF of the relative density $\rho/\langle\rho\rangle = 1 + \delta$ at redshift $z=8$. The RadHydro matter and gas distributions are in very good agreement even at the higher resolution. For AMBER, the TSC assignment scheme with interlacing and deconvolution gives the best agreement, and we only show this case to avoid overcrowding the plot. The results are also in very good agreement at the fiducial resolution, but not at the higher resolution. LPT calculated at second order or even higher order cannot capture the nonlinear shell-crossing on smaller scales and therefore underresolves high-density, collapsed regions. The disagreement in underdense regions is not concerning because it is due to the much smaller number of LPT particles and the different assignment and deconvolution process for AMBER. We would find the same effects for RadHydro if we use a lower-resolution simulation and not done the simple binning.

Figure \ref{fig:rho_cc} shows the density autopower spectra $\Delta_\mathrm{dd}^2(k)$ at redshift $z=8$ and cross correlations relative to the RadHydro matter overdensity $\delta_\mathrm{m}$. All of the dimensionless power spectra continue to increase in amplitude toward smaller scales, typical for cold dark matter dominated models. For the fiducial lower resolution, the power reaches a maximum value of only $\Delta_\mathrm{dd}^2 \approx 0.2$. As this is still in the quasi-linear regime, 2LPT is an accurate and efficient approach for modeling the mass distribution and density field.

The RadHydro matter and gas perfectly trace each other with bias $b_\mathrm{dd} = 1$ and cross correlation $r_\mathrm{dd} = 1$ for $k \lesssim 10\ h/\mathrm{Mpc}$. The suppression in the gas bias on smaller scales is due to the Jeans smoothing from the gas pressure that opposes gravitational collapse. For AMBER, the LPT bias starts out at unity on the largest scales, is still within a few percent for $k \gtrsim 1\ h/\mathrm{Mpc}$, but drops to $\sim 0.9$ by $k \sim 10\ h/\mathrm{Mpc}$. However, the normalized cross correlation is still closer to unity down to smaller scales, showing that the density perturbations are highly in phased even though the amplitudes may differ. We find that using the TSC assignment scheme with interlacing and deconvolution gives more accurate autopower and cross-power spectra. The NGP and CIC results tend to be overbiased and undercorrelated near the Nyquist frequency $k_\mathrm{Ny} = \pi/\Delta x$ because of aliasing. Our results are similar to the previous findings by \citet{2016MNRAS.460.3624S}, who present a thorough analysis of how interlacing can reduce aliasing and improve power spectrum estimation.

\subsubsection{Velocity Fields} \label{sec:velocity_amber}

\begin{figure*}[t]
\center
\includegraphics[width=0.9\textwidth]{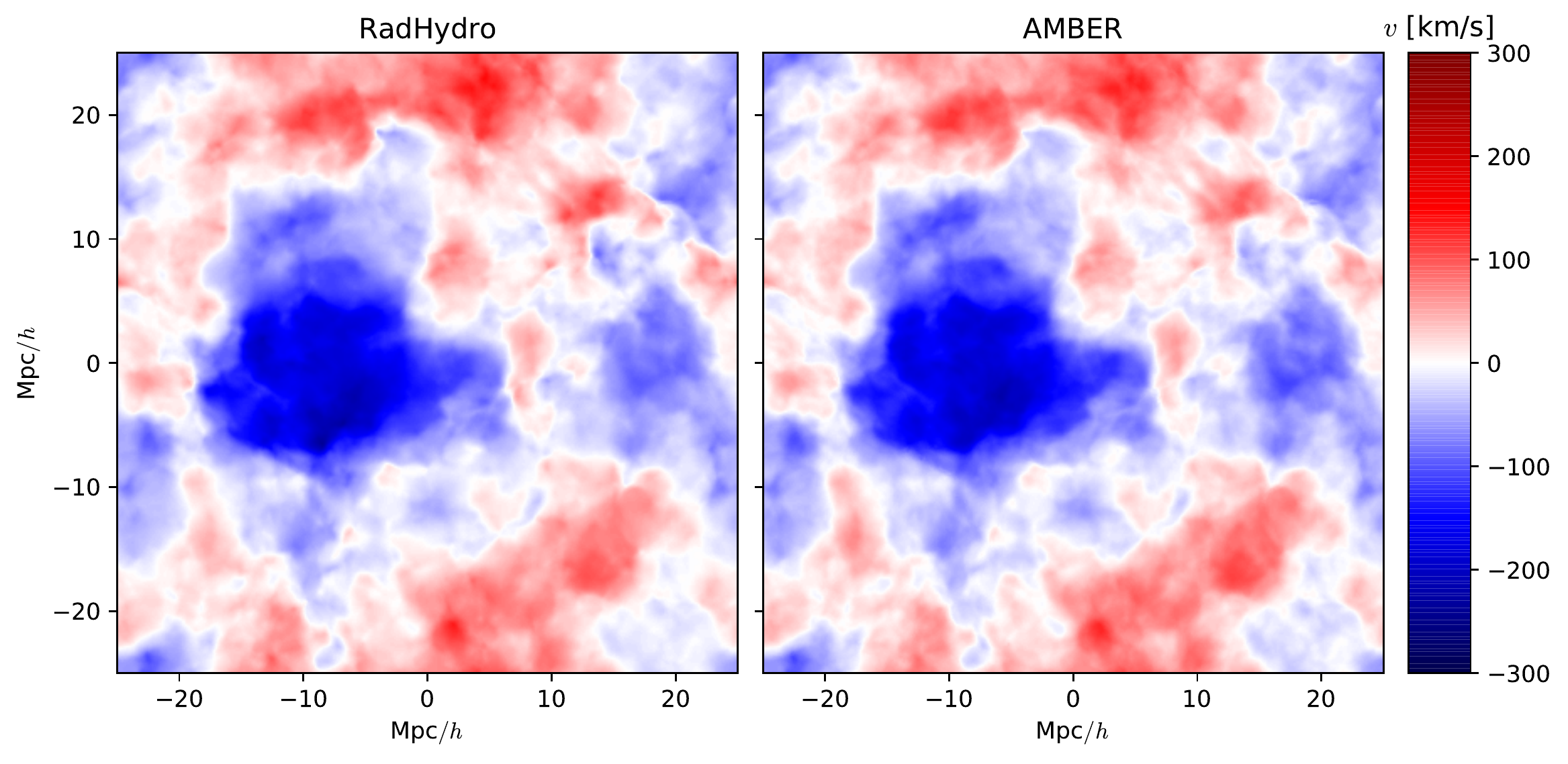}
\caption{Visualization of the line-of-sight component of the velocity fields $\boldsymbol{v}(\boldsymbol{x})$ at redshift $z=8$ from RadHydro (left) and AMBER (right). Each image is $512 \times 512$ pixels from a slice that is $50\, h^{-1}\mathrm{Mpc} \times 50\, h^{-1}\mathrm{Mpc}$ with a thickness of $\sim 1\, h^{-1}\mathrm{Mpc}$. The RadHydro and AMBER velocity fields both show large coherence lengths that are actually underestimated by the moderate simulation box length.}
\label{fig:vel_image}
\end{figure*}

\begin{figure}[t]
\includegraphics[width=\linewidth]{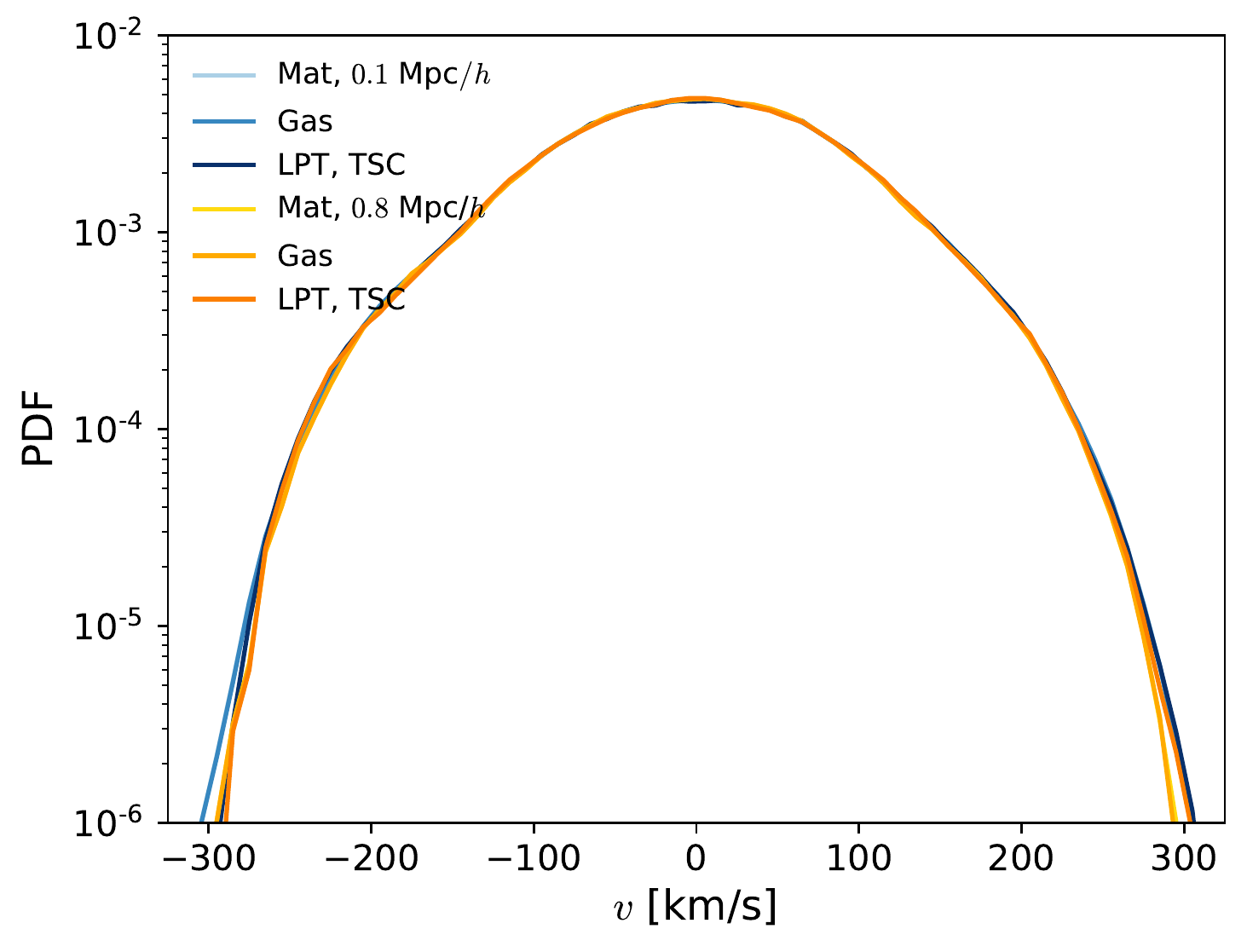}
\caption{One-point probability distribution functions of the velocity components at redshift $z=8$ from RadHydro and AMBER. The RadHydro matter and gas distributions are nearly identical Gaussians, while the AMBER LPT results are also in very good agreement at both resolutions. The differences in the tails of the PDFs are due to nonpresent virialization and imperfect deconvolution for LPT.}
\label{fig:vel_pdf}
\end{figure}

\begin{figure}[t]
\includegraphics[width=\linewidth]{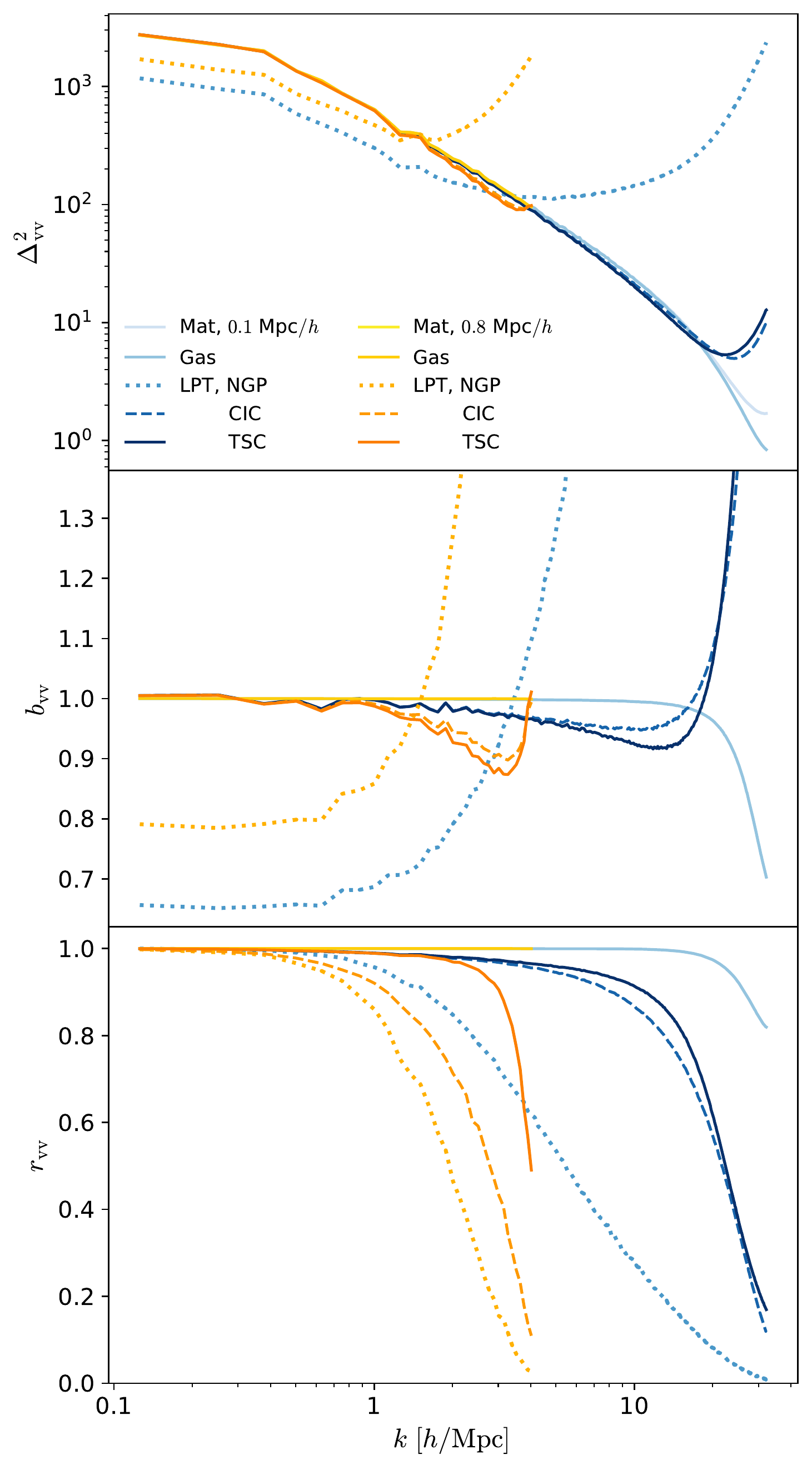}
\caption{{\bf Top:} autopower spectra of the velocity fields at redshift $z=8$ from RadHydro and AMBER. {\bf Center:} for RadHydro, the velocity bias $b_\mathrm{vv}$ of the gas relative to the matter is unity for $k \lesssim 20\ h/\mathrm{Mpc}$. For AMBER, the CIC and TSC assignment schemes with interlacing and deconvolution give similarly accurate bias, but the NGP scheme can produce undefined velocities that lead to incorrect bias. {\bf Bottom:} the cross correlation $r_\mathrm{vv}$ of the gas relative to the matter is unity nearly down to the smallest scale probed. Assigning LPT particles with the TSC scheme significantly improves the cross correlations.}
\label{fig:vel_cc}
\end{figure}

Velocity fields are constructed similarly to the density fields by assigning the LPT particles to interlaced meshes. For a given mesh cell, the velocity $\boldsymbol{v}$ is a weighted average of the individual velocities $\boldsymbol{v}_i$ from all overlapping particles,
\begin{equation} \label{eqn:velocity}
    \boldsymbol{v} = \sum_i \frac{w_i}{w_\mathrm{tot}} \boldsymbol{v}_i = \frac{\sum w_i\boldsymbol{v}_i}{\sum w_i} ,
\end{equation}
where $w_i$ and $w_\mathrm{tot}$ are the individual and total mass assignment weights, respectively. With the NGP assignment scheme, we find that some cells can have no overlapping particles and therefore undefined velocities. Both the CIC and TSC schemes work better in practice with no empty cells and missing velocities for our moderately clustered distribution of LPT particles.

Figure \ref{fig:vel_image} is a visualization of the line-of-sight component of the velocity fields $v(\boldsymbol{x})$ at redshift $z=8$ from RadHydro and AMBER. The velocity images are constructed using the same approach as for the density images (Fig.~\ref{fig:rho_image}). The shown images with the higher-resolution pixels are visually similar and have an even closer resemblance with lower-resolution pixels. The RadHydro and AMBER velocity fields show large coherence lengths that are actually underestimated because of the moderate simulation box length $L = 50\ h^{-1}\text{Mpc}$. We will discuss this interesting feature in more detail below.

Figure \ref{fig:vel_pdf} compares the one-point PDF of the velocity components $v \subset (v_\mathrm{x}, v_\mathrm{y}, v_\mathrm{z})$, at redshift $z=8$. The RadHydro matter and gas velocity distributions are in excellent agreement, even more so than their density distributions. All of the distributions are Gaussian, and the widths decrease slightly by a few percent for the fiducial lower resolution. For AMBER, the CIC and TSC assignment schemes with interlacing and deconvolution give similarly accurate results, and we only show the latter to avoid overcrowding the plot. The differences in the tails of the PDFs are due to the RadHydro simulations having slightly faster velocities because of virialization in rare, collapsed regions.

Figure \ref{fig:vel_cc} shows the velocity autopower spectra $\Delta_\mathrm{vv}(k)$ and the cross correlations relative to the RadHydro matter velocity $v_\mathrm{m}$. We individually compare and average over each velocity component $v \subset (v_\mathrm{x}, v_\mathrm{y}, v_\mathrm{z})$. The velocity power spectrum declines in amplitude toward smaller scales, unlike the density power spectrum (Fig.~\ref{fig:rho_cc}). In linear theory, the transformed velocity field is related to the matter overdensity field as $\boldsymbol{v}(\boldsymbol{k}) \propto (\boldsymbol{k}/k^2)\delta(\boldsymbol{k})$, and therefore the power is predominantly coming from larger scales. This explains the large coherence length seen in the velocity images (Fig.~\ref{fig:vel_image}). While radiation-hydrodynamic simulations are still computationally too expensive, semi-numerical methods are sufficiently efficient to afford large box sizes of $\gtrsim 500\ h^{-1}\text{Mpc}$, which is necessary to capture the large-scale power and accurately model the velocity fields.

The RadHydro matter and gas velocities perfectly trace each other with bias $b_\mathrm{vv} = 1$ and cross correlation $r_\mathrm{vv} = 1$ for $k \lesssim 20\ h/\mathrm{Mpc}$, even to smaller scales compared to their densities (Fig.~\ref{fig:rho_cc}). For AMBER, the NGP assignment scheme gives incorrect bias and poor stochasticity when we simply set the velocities to zero where ever it is undefined. The CIC and TSC assignment schemes have similarly accurate bias, but the latter has preferably better cross correlation. For the velocity bias, the lower-resolution results diverge from the higher-resolution ones at larger scales compared to the density bias. The velocity field is a weighted average of the particle velocities (Eq.~\ref{eqn:velocity}) and a simple deconvolution using the mass assignment window function (Eq.~\ref{eqn:Wofk}) is only approximately correct.

In AMBER, we adopt the TSC assignment scheme with interlacing and deconvolution as the default method for constructing density and velocity fields from the LPT particles. This approach reduces aliasing and smoothing, and produces the most accurate one-point distributions and two-point correlations. Accurate density and velocity fields are essential components for modeling EoR observables. The 21cm signal and Ly$\alpha$ forest both depend on the neutral hydrogen density $n_\mathrm{HI}(\boldsymbol{x})$ and line-of-sight velocity field $v_\mathrm{los}(\boldsymbol{x})$. For the CMB, the patchy Thomson optical depth and patchy KSZ effect depend on the electron number density $n_\mathrm{e}(\boldsymbol{x})$ and line-of-sight momentum $n_\mathrm{e}(\boldsymbol{x})v_\mathrm{los}(\boldsymbol{x})$, respectively.

\subsection{Excursion Set Formalism} \label{sec:esf_amber}

\begin{figure*}[t]
\includegraphics[width=\textwidth]{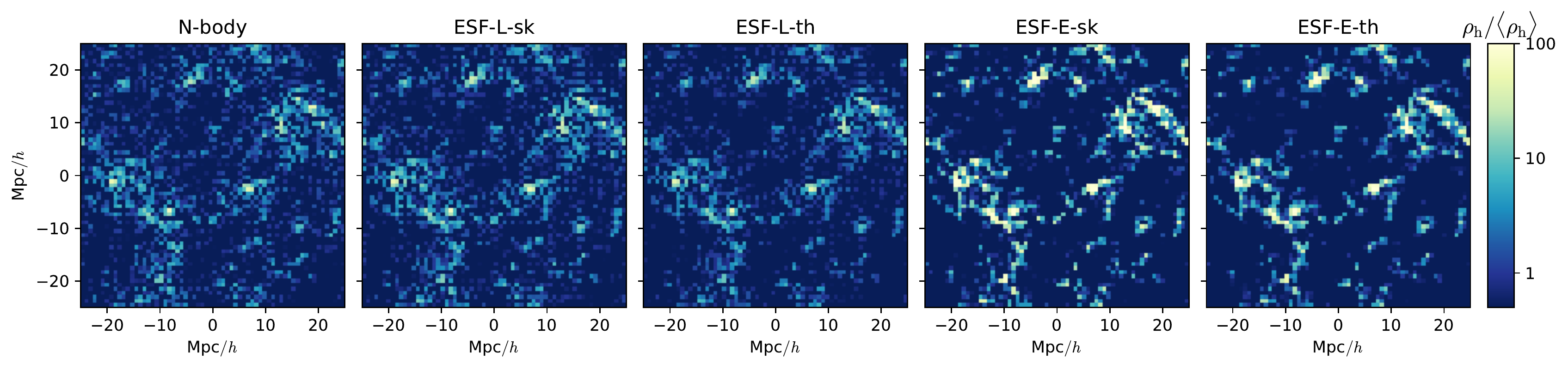}
\caption{Visualization of the halo mass density fields $\rho_\mathrm{h}(\boldsymbol{x})/\langle \rho_\mathrm{h} \rangle$ for $M_\mathrm{min} = 10^8\ h^{-1}M_\odot$ at $z=8$ from a N-body simulation and with the Lagrangian (L) and Eulerian (E) implementations of the excursion set formalism (ESF). For ESF-L and ESF-E, we study both sharp $k$-space (sk) and tophat (th) filters. Each image is $64 \times 64$ pixels from a slice that is $50\, h^{-1}\mathrm{Mpc} \times 50\, h^{-1}\mathrm{Mpc}$ with a thickness of $\sim 1\, h^{-1}\mathrm{Mpc}$. The ESF-L images appear very similar to the N-body result, but the ESF-E images show higher halo mass densities in collapsed regions.}
\label{fig:halo_image}
\end{figure*}

The ESF \citep{1991ApJ...379..440B} is used to model the collapsed mass fraction and the halo mass function in extended Press-Schecter theory (EPS). ESF has the advantage of being orders of magnitude faster than an N-body simulation and halo finding, but it is known to only produce approximately correct results. In the majority of semi-numerical methods for reionization, it is used to model both the collapsed mass and ionization fraction field \citep[e.g.][]{2004ApJ...613....1F}. For AMBER, we use ESF only for modeling the halo mass density field.

We will first summarize ESF and then test its accuracy for modeling the halo mass density field in AMBER. The linearly extrapolated overdensity field $\delta_0$ is filtered to produce
\begin{align}
    \delta_\mathrm{f}(\boldsymbol{x}) & = \int\delta_0(\boldsymbol{x}^\prime)W_\mathrm{f}(\boldsymbol{x} - \boldsymbol{x}^\prime) d^3x^\prime , \\
    \delta_\mathrm{f}(\boldsymbol{k}) & = \delta_0(\boldsymbol{k})W_\mathrm{f}(\boldsymbol{k}) ,
\end{align}
where the window function $W_\mathrm{f}$ is usually taken to be a sharp $k$-space filter,
\begin{align}
    W_\mathrm{sk}(r) & = \frac{1}{2\pi^2 r^2}\left[\sin(r/R_\mathrm{sk}) - (r/R_\mathrm{sk})\cos(r/R_\mathrm{sk})\right] , \\
    W_\mathrm{sk}(k) & = \begin{cases}
    1 \quad & kR_\mathrm{sk} \leq 1 \\
    0 & \text{otherwise}
    \end{cases} ,
\end{align}
or a spherical tophat filter,
\begin{align}
    W_\mathrm{th}(r) & = 
    \begin{cases}
    3/(4\pi R_\mathrm{th}^3) \quad & r \leq R_\mathrm{th} \\
    0 & \text{otherwise}
    \end{cases} , \\
    W_\mathrm{th}(k) & = \frac{3}{(kR_\mathrm{th})^3}\left[\sin(kR_\mathrm{th}) - (kR_\mathrm{th})\cos(kR_\mathrm{th})\right] .
\end{align}
The convolution is rapidly computed in Fourier space after transforming with highly optimized FFTs.

In a large-scale region with filtered overdensity $\delta_\mathrm{f}$, the collapsed fraction of matter in halos above a minimum mass $M_\mathrm{min}$ is given by the EPS result \citep{1993MNRAS.262..627L},
\begin{equation}
    \label{eqn:fcoll}
    f_\mathrm{coll} = \mathrm{erfc}\left[\frac{\delta_\mathrm{c}(z) - \delta_\mathrm{f}}{\sqrt{2[\sigma^2(M_\mathrm{min}) - \sigma^2(M_\mathrm{s})]}}\right] .
\end{equation}
The variance of the linear density fluctuations, smoothed on mass scale $M_\mathrm{s}$, is calculated as
\begin{equation}
    \label{eqn:variance}
    \sigma^2(M_\mathrm{s}) = \int \Delta^2_\mathrm{lin}(k)|W_\mathrm{th}(k)|^2 d\mathrm{ln}k ,
\end{equation}
where $\Delta^2_\mathrm{lin}(k)$ is the dimensionless linear power spectrum, and $W_\mathrm{th}(k)$ is the spherical tophat filter with comoving radius $R_\mathrm{s} = [M_\mathrm{s}/(4\pi\bar{\rho}_0/3)]^{1/3}$. The corresponding collapsed overdensity barrier is given by
\begin{equation} \label{eqn:deltac}
    \delta_\mathrm{c}(z) = \frac{ \delta_\mathrm{crit}}{D(z)} ,
\end{equation}
where $\delta_\mathrm{crit} \approx 1.686$ is the spherical collapse value, and $D(z)$ is the linear growth factor. Note that the redshift dependence only explicitly shows up in Equation \ref{eqn:deltac}, while the other functions and fields have already been linearly extrapolated to $z=0$.

In the original version of ESF \citep[e.g.][]{1991ApJ...379..440B, 1993MNRAS.262..627L}, the filtering is done in Lagrangian ($\boldsymbol{q}$) space on the initial overdensity field. In an alternative version that is used in some semi-numerical models of reionization \citep[e.g][]{2011MNRAS.411..955M, 2011MNRAS.414..727Z}, the filtering is done in Eulerian ($\boldsymbol{x}$) space on the evolved density field. We will refer to these versions as ESF-L and ESF-E, respectively.

When the filter window function $W_\mathrm{f}$ is chosen to be a spherical tophat function, it is natural to set the filter radius $R_\mathrm{th}$ to be equal to the comoving radius $R_\mathrm{s}$ in the variance. For the sharp $k$-space function, it is unclear how to choose the filter radius $R_\mathrm{sk}$. Following \citet{1993MNRAS.262..627L}, we will simply choose $R_\mathrm{sk} = [M_\mathrm{s}/(6\pi\bar{\rho}_0)]^{1/3} =  [2/(9\pi)]^{1/3}R_\mathrm{s}$.

For ellipsoidal collapse, \citet*{2001MNRAS.323....1S} propose a moving barrier $\delta_\mathrm{c}(\sigma,z)$ with three coefficients that can be changed in order to match the halo mass functions from N-body simulations. Their moving barrier is used for determining when halos collapse following the peak-patch approach of \citet{1996ApJS..103....1B}. However, it is not self-consistent to simply use their $\delta_\mathrm{c}(\sigma,z)$ in Equation \ref{eqn:fcoll} to calculate the collapsed fraction. Furthermore, the proposed functional form for $\delta_\mathrm{c}(\sigma,z)$ may not be valid for all halo mass functions \citep[e.g.][]{2009ApJ...696..636R}. The \citet{1999MNRAS.308..119S} halo mass function corresponding to this moving barrier does not agree with the more recent halo mass functions for the EoR in SCORCH I \citep{2015ApJ...813...54T}. Therefore, we will use the spherical overdensity $\delta_\mathrm{c}(z)$ rather than the ellipsoidal collapse barrier.

\subsubsection{Halo Mass Density Fields} \label{sec:halo_amber}

\begin{figure}[t]
\includegraphics[width=\linewidth]{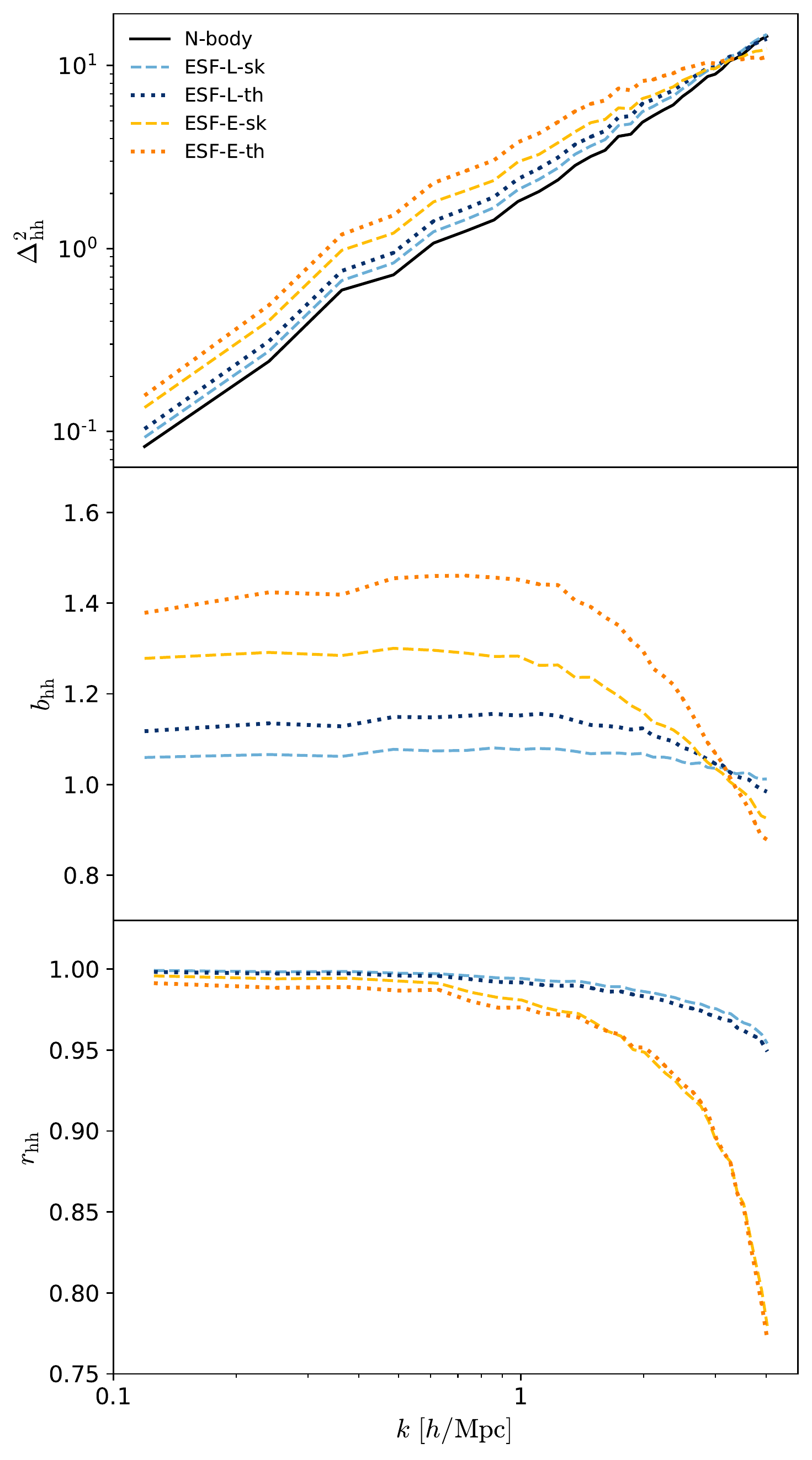}
\caption{{\bf Top:} autopower spectra of the halo mass density fields for $M_\mathrm{min} = 10^8\ h^{-1}M_\odot$ at $z=8$ from an N-body simulation and from ESF-L and ESF-E with sharp $k$-space (sk) and tophat (th) filters. {\bf Center:} relative to the N-body halos, ESF-L has better halo bias $b_\mathrm{hh}$ that is closer to unity than ESF-E. {\bf Bottom:} ESF-L also has better halo cross correlation $r_\mathrm{hh}$ that is closer to unity than ESF-E.}
\label{fig:halo_cc}
\end{figure}

\begin{figure*}[t]
\includegraphics[width=\textwidth]{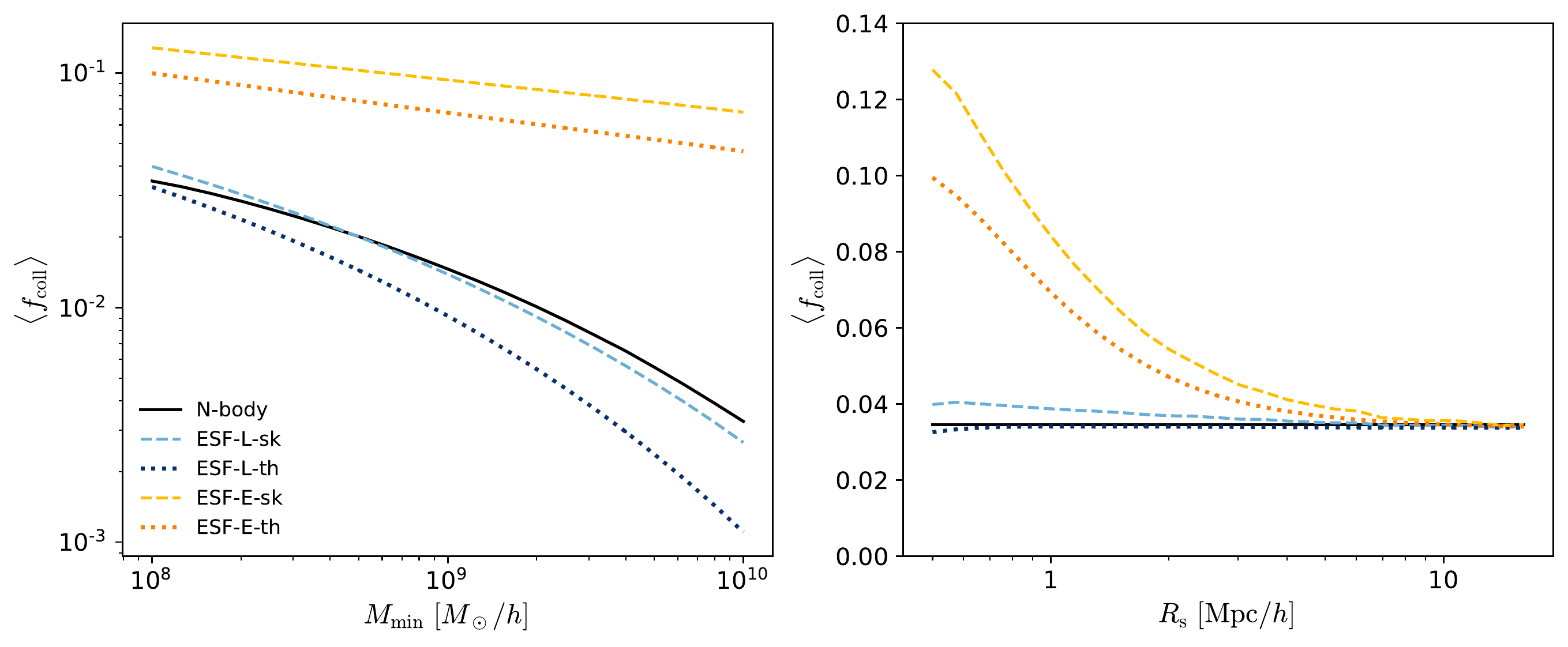}
\caption{{\bf Left:} the mass-weighted average collapse fraction of matter in halos above a minimum mass $M_\mathrm{min}$ at fixed ESF smoothing radius $R_\mathrm{s} = 0.5\ h^{-1}\mathrm{Mpc}$. ESF-L with the sharp $k$-space filter gives the best agreement with the N-body halo results, while ESF-E significantly overpredicts the average collapse fraction. {\bf Right:} the variation in the average collapse fraction with $R_\mathrm{s}$ at 
fixed $M_\mathrm{min} = 10^8\ h^{-1}M_\odot$. The ESF-L values only vary by $\lesssim 10\%$, while the ESF-E results are highly sensitive to the choice of smoothing and filtering scales.}
\label{fig:halo_fcoll}
\end{figure*}

We construct halo mass density fields with ESF and compare them against results from a high-resolution N-body simulation. For reference, we use the halo catalog from SCORCH II (Sec.~\ref{sec:nbody_scorch}) and select all halos above a minimum mass $M_\mathrm{min} = 10^8\ h^{-1}M_\odot$, which corresponds to the atomic cooling limit for galaxy formation at redshift $z \approx 8$. To assign the halo mass to interlaced meshes, we choose the CIC scheme as it gives both reduced aliasing and strong correlation with the matter density field. The NGP scheme has the most aliasing, while the TSC scheme has the most stochasticity as it spreads the small halos over too many larger grid cells.

For ESF-L, we calculate the
collapsed fraction and mass in Lagrangian space,
\begin{equation}
    m_\mathrm{coll}(\boldsymbol{q}) = f_\mathrm{coll}(\boldsymbol{q})m_\mathrm{p} ,
\end{equation}
for LPT particles of mass $m_\mathrm{p}$, move the particles to their comoving positions $\boldsymbol{x}$ (Eq.~\ref{eqn:xlpt}), and then assign the collapsed mass to the interlaced meshes (Eq.~\ref{eqn:interlace}) to obtain the halo mass density field $\rho_\mathrm{h}(\boldsymbol{x})$. For ESF-E, we calculate the collapsed fraction field in Eulerian space, and then the halo mass density field is calculated as
\begin{equation}
    \rho_\mathrm{h}(\boldsymbol{x}) = f_\mathrm{coll}(\boldsymbol{x})\bar{\rho}_\mathrm{m}[1 + D(z)\delta_\mathrm{f}(\boldsymbol{x})] ,
\end{equation}
where $\bar{\rho}_\mathrm{m}$ is the average matter density, and the growth factor $D(z)$ reverses the linear extrapolation done earlier. 

ESF requires that the mass smoothing scale $M_\mathrm{s}$ be greater than the minimum halo mass $M_\mathrm{min}$ in Equation \ref{eqn:fcoll}. The smoothing scale also cannot be lower than the particle mass resolution. For the fiducial lower resolution with $64^3$ particles on a $64^3$ mesh in the $50\ h^{-1}\mathrm{Mpc}$ box, the particle mass is $m_\mathrm{p} = 4.0 \times 10^{10}\ h^{-1}M_\odot$ and the cubical cell size is $l_\mathrm{c} = 0.8\ h^{-1}\mathrm{Mpc}$. Therefore, we will choose and vary the smoothing scales such that $M_\mathrm{s} \ge m_\mathrm{p}$ and $R_\mathrm{s} = [M_\mathrm{s}/(4\pi\bar{\rho}_0/3)]^{1/3} \gtrsim 0.5\ h^{-1}\mathrm{Mpc}$.

Figure \ref{fig:halo_image} is a visualization of the halo mass density fields $\rho_\mathrm{h}(\boldsymbol{x})/\langle \rho_\mathrm{h} \rangle$ for $M_\mathrm{min} = 10^8\ h^{-1}M_\odot$ at $z=8$. Here, we normalize each density field using the same mean $\langle \rho_\mathrm{h} \rangle$ taken from the N-body result to preserve the relative differences in $\rho_\mathrm{h}(\boldsymbol{x})$. We also use the fiducial smoothing scale $M_\mathrm{s} = m_\mathrm{p}$. The halo images appear more pixelated than the corresponding density and velocity images (Fig.~\ref{fig:rho_image},\ref{fig:vel_image}) because of the lower resolution shown here. The ESF-L images with the sharp $k$-space and tophat filters are remarkably similar to the N-body result, but the ESF-E images are noticeably different. ESF-E produces relatively higher halo densities in prominent collapsed regions and lower values elsewhere, and this larger range of density contrast can be traced back to using the evolved density field rather than the initial density field for the filtering process.

Figure \ref{fig:halo_cc} shows the halo autopower spectra $\Delta_\mathrm{hh}^2(k)$ and cross correlations. The halo power spectrum is approximately power-law with slope $n = d\ln P/d\ln k \approx -1.5$ at $z=8$. In comparison with the matter power spectrum (Fig.~\ref{fig:rho_cc}), we find scale-dependent halo-matter bias $b_\mathrm{hm}(k)$ with a large-scale value of $\approx 3.7$. Similarly, the halo-matter cross correlation $r_\mathrm{hm}(k)$ is close to unity on the largest scales, but drops below 0.9 for $k \gtrsim 1\ h/\mathrm{Mpc}$. Therefore, we cannot accurately model the halo mass density field assuming linear bias, even if we allow for scale dependence.

In cross correlation with the N-body halo mass density field, the ESF-L with the sharp $k$-space filter gives the best agreement, closely followed by with the tophat filter. Both the bias $b_\mathrm{hh}(k)$ and cross correlation $r_\mathrm{hh}(k)$ differ from unity by $\lesssim 0.05$ down to the smallest scale shown. The strong agreement is remarkable given that we have adopted nominal choices for the filter radius $R_\mathrm{sk}$ and collapse barrier $\delta_\mathrm{c}$ without any fine-tuning. ESF-E with either the sharp $k$-space or tophat filter gives poorer agreement with the N-body results. The larger amplitude in the power spectrum and bias is consistent with the larger range in the density contrast seen in the halo density images (Fig.~\ref{fig:halo_image}).

We now vary the minimum halo mass $M_\mathrm{min}$ and smoothing mass $M_\mathrm{s}$ and quantify the mass-weighted average collapse fraction $\langle f_\mathrm{coll}\rangle$ in Figure \ref{fig:halo_fcoll}. As $M_\mathrm{min}$ is increased above the atomic cooling limit while fixing $M_\mathrm{s} = 4 \times 10^{10}\ h^{-1}M_\odot$ and $R_\mathrm{s} = 0.5\ h^{-1}\mathrm{Mpc}$, we again find that ESF-L with the sharp $k$-space filter gives the best agreement, followed by with the tophat filter. However, ESF-E overpredicts the average collapse fraction by a factor of $\gtrsim 3$, with increasingly larger differences toward higher $M_\mathrm{min}$. Note that we should expect small differences in the average collapse fraction. In the N-body halo finder \citep{2015ApJ...813...54T}, the halo mass $M_{200}$ is defined such that the average halo density $\bar{\rho}_\mathrm{h}$ is 200 times the average matter density  $\bar{\rho}_\mathrm{m}(z)$, which is also equal the critical density $\bar{\rho}_\mathrm{crit}(z)$ at high redshifts. In ESF, the halo mass is not clearly defined, although in spherical collapse theory, the average halo density is usually $18\pi^2 \approx 180$ times the critical density at high redshifts or in Einstein-de Sitter cosmologies.

As $M_\mathrm{s}$ and $R_\mathrm{s}$ are increased while fixing $M_\mathrm{min} = 10^8\ h^{-1}M_\odot$, we find that ESF-L gives reliably accurate and stable values for the average collapsed fraction that are within 10\% of the N-body halo value. However, ESF-E gives highly variable results that are only in agreement at large smoothing radius $R_\mathrm{s} \gtrsim 10\ h^{-1}\mathrm{Mpc}$. Similarly, we also find that the power and bias change as the smoothing is varied, which is not found for ESF-L. 

The sensitivity of ESF-E to the smoothing and filtering scales is especially troublesome for the semi-numerical models of reionization that are based on this approach. For a given ionized region, the filtering scale is adaptively varied until the number of ionizing photons equals the number of hydrogen atoms. For the entire modeled volume, there is a distribution of filtering scales rather than just one value, and as a consequence, there is no one value for the average collapse fraction and large-scale halo bias. Furthermore, as one changes the reionization parameters, the distribution of filtering scales will change, and so too will the halo mass density field and power spectrum, both of which should be independent of the reionization history.

In AMBER, we adopt the ESF-L with the sharp $k$-space filter as the default choice for modeling the halo mass density field. In the next section, we will discuss that our abundance matching technique does not depend on the overall average density $\langle \rho_\mathrm{h}\rangle$ nor the overall normalization of the halo bias $b_\mathrm{hm}$. In future work, we can calibrate the EPS collapsed fraction relation (Eq.~\ref{eqn:fcoll}) using the N-body simulations and halo catalogs from SCORCH I and II. Accurate halo abundance and density fields are required to properly model high-redshift galaxies and quasars as radiation sources.

\subsection{Abundance Matching} \label{sec:abundmatch_amber}

The new idea for AMBER is that there is a spatial order to the reionization process, and abundance matching can be applied to a correlated field to accurately predict the reionization-redshift field (Sec.~\ref{sec:zre_scorch}) such that the reionization history follows a given mass-weighted ionization fraction (Sec.~\ref{sec:reionhistory_amber}). We emphasize that the abundance matching technique does not depend on the overall normalization of the correlated field of interest. In \citet{2008ApJ...689L..81T}, we find that the higher-density regions near sources are generally reionized earlier than the lower-density regions far from sources in our RadHydro simulations. Following up in \citet{2013ApJ...776...81B}, we find that the density and reionization-redshift fields are correlated down to Mpc scales. While two natural choices for abundance matching are the matter density field (Sec.~\ref{sec:density_amber}) and the halo mass density field (Sec.~\ref{sec:halo_amber}), the radiation field is more directly relevant to the photoionization of hydrogen.

\subsubsection{Radiation Field} \label{sec:radfield_amber}

The radiation field is specified by the hydrogen photoionization rate,
\begin{equation} \label{eqn:gammaHI}
    \Gamma_\mathrm{HI}(\boldsymbol{x}) = 4\pi \int_{\nu_\mathrm{HI}}^\infty \frac{J_\nu(\nu,\boldsymbol{x})}{h\nu}\sigma_\mathrm{HI}(\nu)d\nu ,
\end{equation}
where $J_\nu$ is the specific intensity, $\sigma_\mathrm{HI}$ is the hydrogen photoionization cross-section, and $\nu_\mathrm{HI}$ is the Lyman limit frequency. In our RadHydro simulations, the RT is calculated using an accurate but expensive raytracing algorithm \citep{2007ApJ...671....1T}. However, this is an unnecessarily costly approach for calculating the radiation field just for abundance matching, which does not depend on the overall normalization. In AMBER, we model the specific intensity field as
\begin{equation} \label{eqn:intensity}
    J_\nu(\nu,\boldsymbol{x}) = \int \frac{S_\nu(\nu,\boldsymbol{x}^\prime)}{4\pi|\boldsymbol{x} - \boldsymbol{x}^\prime|^2}\exp\left(-\frac{|\boldsymbol{x} - \boldsymbol{x}^\prime|}{l_\mathrm{mfp}}\right) d^3x^\prime ,
\end{equation}
where $S_\nu$ is the source function, the $1/(4\pi r^{2})$ term is the inverse-square law for flux, and the $e^{-r/l_\mathrm{mfp}}$ term is the transmitted fraction over the radiation mean free path $l_\mathrm{mfp}$.

The source field is related to the specific luminosity density and can be expressed as
\begin{equation}
    S_\nu(\nu,\boldsymbol{x}) = f_\mathrm{esc}\epsilon_\nu(\nu)\rho_\mathrm{sfr}(\boldsymbol{x}) ,
\end{equation}
where $\rho_\mathrm{sfr}$ is the star formation rate density, $\epsilon_\nu$ is the radiative energy per unit star formation rate per frequency, and $f_\mathrm{esc}$ is the radiation escape fraction \citep[e.g.][]{1999ApJ...514..648M, 2010Natur.468...49R}. The star formation rate density field is often modeled using the halo mass density field,
\begin{equation}
    \rho_\mathrm{sfr}(\boldsymbol{x}) = \frac{f_\mathrm{star}\rho_\mathrm{halo}(\boldsymbol{x})}{\tau_\mathrm{star}} ,
\end{equation}
where $f_\mathrm{star}$ and $\tau_\mathrm{star}$ are the star formation efficiency and timescale, respectively \citep[e.g.][]{1992ApJ...399L.113C, 2007ApJ...671....1T}. The source field is proportional to the halo mass density field, while the normalization depends on the product of several astrophysical parameters and functions ($f_\mathrm{esc}$, $f_\mathrm{star}$, $\tau_\mathrm{star}$, $\epsilon_\nu$), which may have redshift dependence that further alters the reionization history. In semi-numerical methods based on the ESF, this combination is often represented by a single efficiency parameter $\zeta$, which is generally assumed to be constant for simplicity. With our abundance matching technique, we do not need to specify the astrophysical quantities to calculate the normalization of the the source field, but instead get to cast them and their effects in terms of the the redshift midpoint, duration, and asymmetry parameters that directly set the reionization history (Sec.~\ref{sec:reionhistory_amber}).

In AMBER, we adopt the common assumption $S_\nu(\boldsymbol{x}) \propto \rho_\mathrm{halo}(\boldsymbol{x})$ and ignore the overall normalization for the source field as it is not necessary for abundance matching. The radiation field is computed quickly using FFTs to perform the convolution in Equation \ref{eqn:intensity}. In practice, we can abundance match the specific intensity field instead of the photoionization rate field, as the integration over frequency only changes the overall normalization for the latter.

We use an effective mean free path to account for the attenuation of the radiation field. Our use of a mean free path parameter is more physical than the maximum bubble size imposed in some semi-numerical models based on the ESF. It is more self-consistent with RT theory \citep[e.g.][]{1999ApJ...514..648M} and in better agreement with observational measurements \citep[e.g.][]{2014MNRAS.445.1745W} extrapolated to higher redshifts for the EoR. It also helps to mitigate uncertainties on radiation sinks. Recently, \citet{2021arXiv210309821D} have also introduced an ionizing photon mean free path as an improved alternative to the maximum filtering scale in ESF methods like 21cmFAST. Their implementation is different in that $l_\mathrm{mfp}$ ultimately shows up in their scale-dependent ionization criterion, whereas we use it to directly compute an attenuated radiation field.

\subsubsection{Reionization-redshift Fields} \label{sec:zream_amber}

\begin{figure*}[t]
\includegraphics[width=\textwidth]{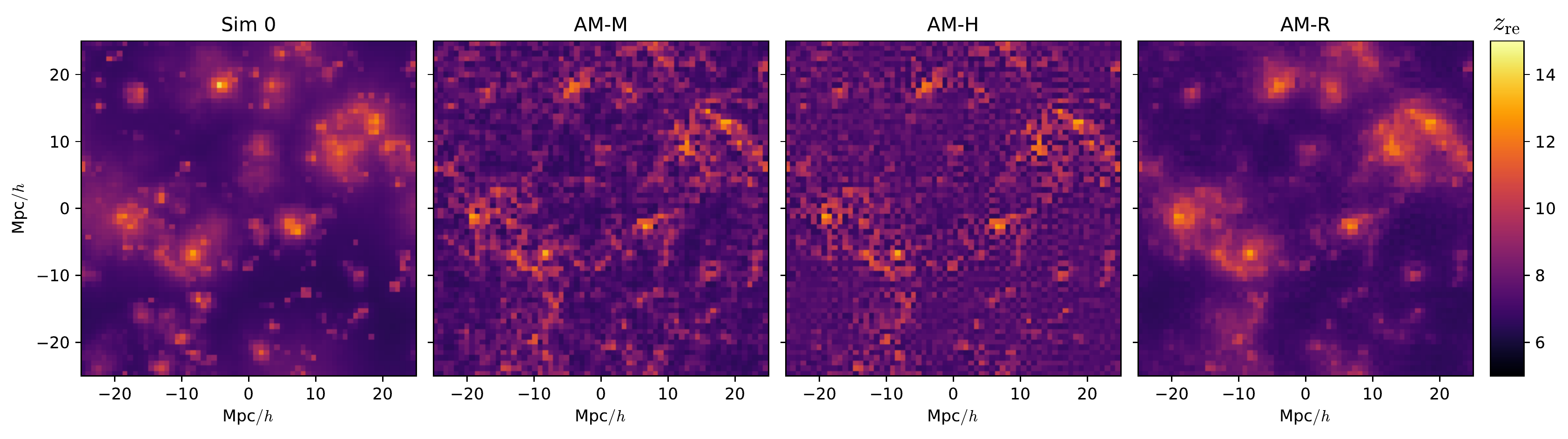}
\includegraphics[width=\textwidth]{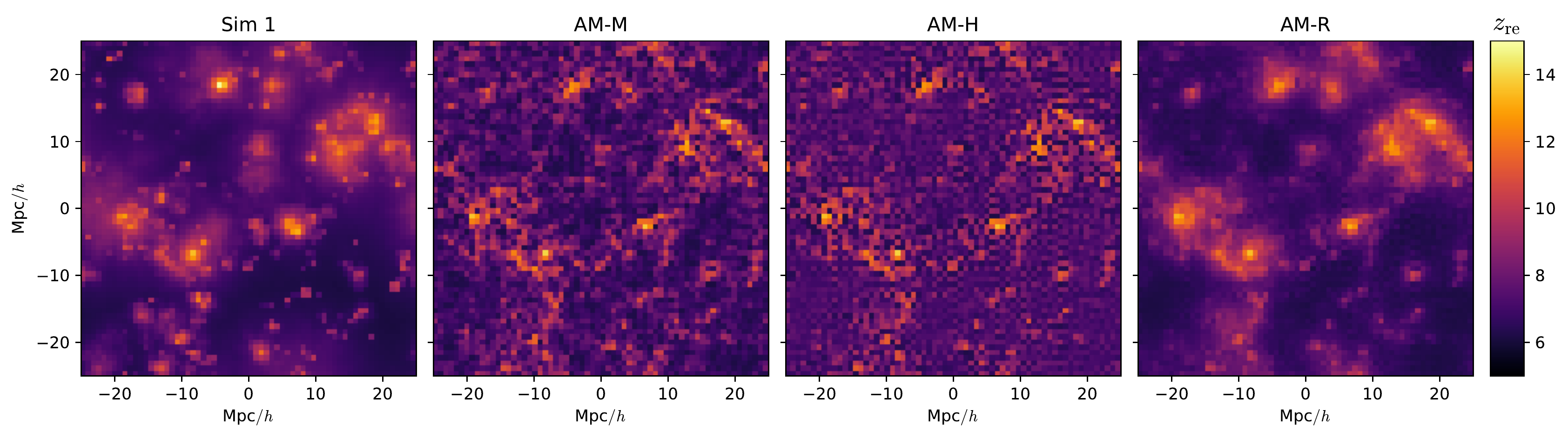}
\includegraphics[width=\textwidth]{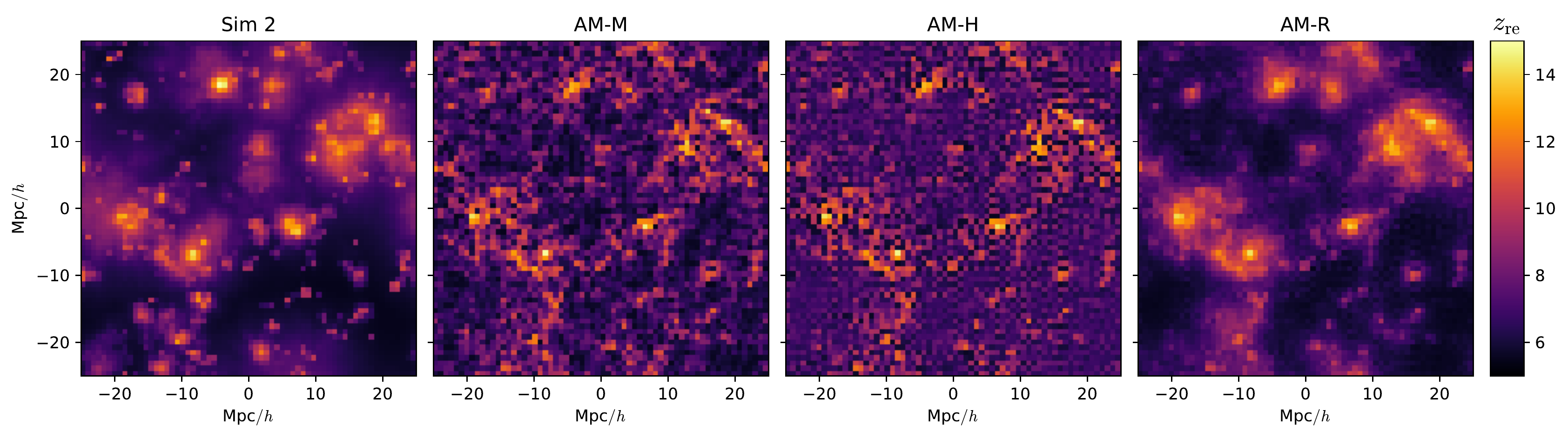}
\caption{Visualization of the reionization-redshift fields $z_\mathrm{re}(\boldsymbol{x})$ from the three RadHydro simulations (left) and corresponding abundance matching models using matter density (AM-M), halo mass density (AM-H), and radiation intensity (AM-R) fields. Each image is $64 \times 64$ pixels from a slice that is $50\, h^{-1}\mathrm{Mpc} \times 50\, h^{-1}\mathrm{Mpc}$ with a thickness of $\sim 1\, h^{-1}\mathrm{Mpc}$. Abundance matching using the radiation field more correctly captures that large-scale regions near sources are generally reionized earlier than large-scale regions far from sources, and is the adopted approach in AMBER.}
\label{fig:zre_image}
\end{figure*}

\begin{figure}[t]
\includegraphics[width=\linewidth]{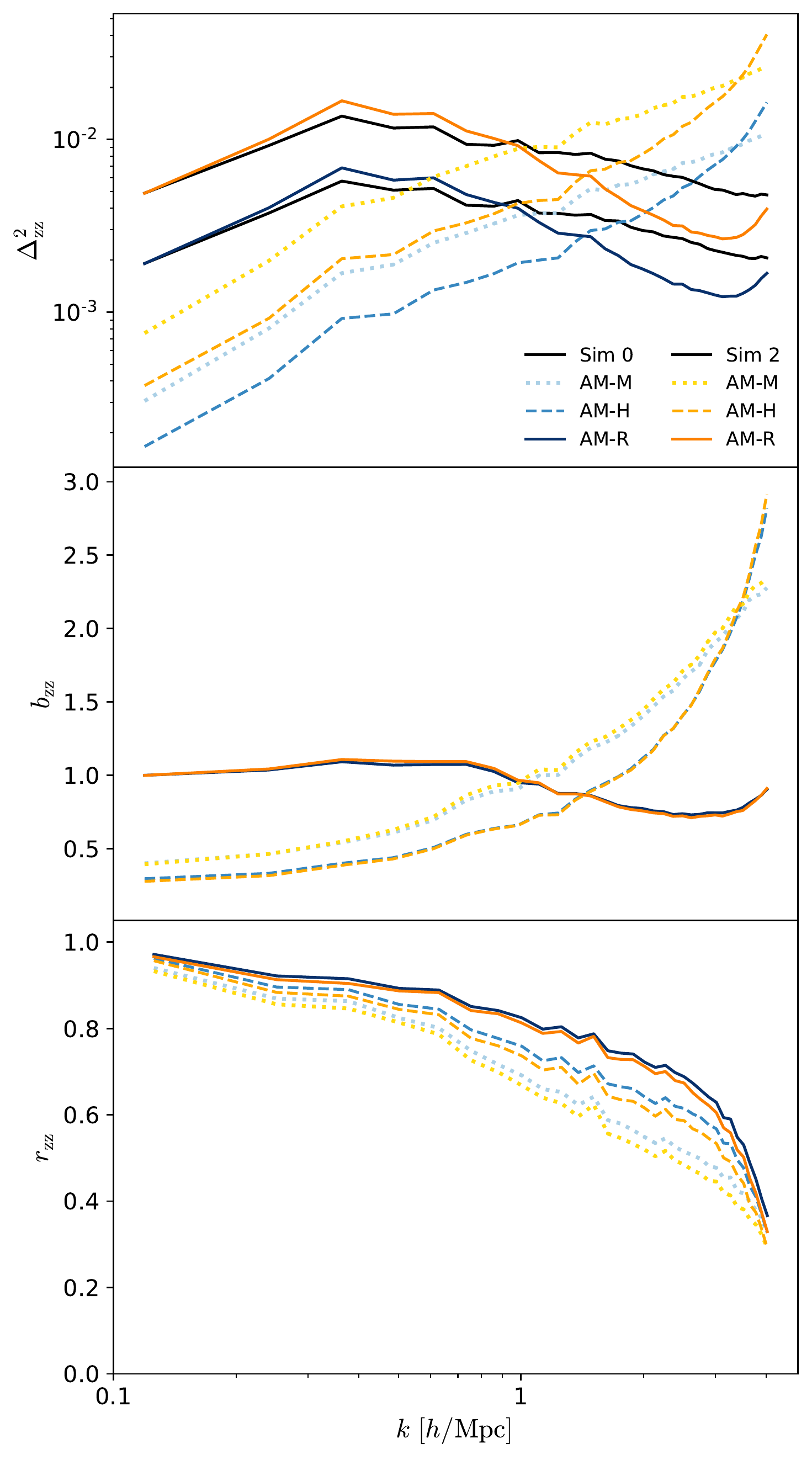}
\caption{{\bf Top:} autopower spectra of the reionization-redshift fields from the RadHydro simulations and corresponding abundance matching models using the matter density (AM-M), halo mass density (AM-H), and radiation intensity (AM-R) fields. The AM-R spectra have a characteristic peak similar to the simulations. {\bf Center:} the redshift bias $b_\mathrm{zz}$ between the models and simulations for AM-R are much closer to unity than for AM-M and AM-H. {\bf Bottom:} the AM-R models have improved cross correlation $r_\mathrm{zz}$ that approach unity on large scales compared to AM-M and AM-H.}
\label{fig:zre_cc}
\end{figure}

The reionization-redshift field $z_\mathrm{re}(\boldsymbol{x})$ is assumed to be correlated with a given field, either the matter density field, halo mass density field, or radiation field. A region with higher matter density, halo density, or radiation intensity is considered to be photoionized earlier and has a higher reionization redshift. The abundance matching technique assigns redshift values such that the reionization history follows a given mass-weighted ionization fraction $\bar{x}_\mathrm{i}(z)$, specified with the redshift midpoint, duration, and asymmetry parameters and interpolated with a Weibull function (Sec.~\ref{sec:reionhistory_amber}).

We perform the abundance matching on a correlated field at a single redshift for computational efficiency, but it can also be done tomographically using multiple redshift intervals. The correlated field is first constructed at the redshift midpoint $z_\mathrm{mid}$ and the data array is ranked in descending order using a parallel quicksort. For the $n$th rank order cell out of a total of $N$, all cells with indices $m\leq n$ are considered ionized. The reionization redshift $z_n$ is calculated by equating the cumulative mass fraction with the mass-weighted ionization fraction:
\begin{equation}
\frac{\sum_{m=1}^{n} 1 + \delta_m(z_n)}{\sum_{m=1}^N 1 + \delta_m(z_n)} = \bar{x}_\text{i}(z_n) ,
\end{equation}
where the matter overdensity for the $m$th cell is linearly extrapolated from the redshift midpoint,
\begin{equation}
    \delta_m(z_n) = \frac{D(z_n)}{D(z_\mathrm{mid})} \delta_m(z_\mathrm{mid}) .
\end{equation}
The linear growth is appropriate for the modest overdensities in lower-resolution semi-numerical models, and because we have chosen the redshift midpoint as the pivot.

We construct the abundance matching models using the matter density field (AM-M), halo mass density field (AM-H), and radiation intensity fields (AM-R), and compare them with the RadHydro reionization-redshift fields (Sec.~\ref{sec:zre_scorch}). For the AM-M models, we use the TSC particle assignment scheme with interlacing and deconvolution to construct the matter density field (Sec.~\ref{sec:density_amber}). For the AM-H models, we use the ESF-L with sharp $k-$space filter for constructing the halo mass density field and set the minimum halo mass to $M_\mathrm{min} = 10^8\ h^{-1}M_\odot$ (Sec.~\ref{sec:halo_amber}). While this corresponds to the threshold for atomic cooling halos, the RadHydro simulations are more complex because of the nonmonotonic star formation efficiency and episodic star formation \citep{2015ApJ...813...54T, 2019ApJ...870...18D}. For the AM-R models, we vary the radiation mean free path to find the best agreement between the reionization-redshift fields.

Figure \ref{fig:zre_image} is a visualization of the reionization-redshift fields $z_\mathrm{re}(\boldsymbol{x})$ from the RadHydro simulations and abundance matching models. While all of the models have the same reionization history for a given simulation, their reionization-redshift fields are visibly different. The AM-R images most closely resemble the simulation results and correctly show that large-scale regions near radiation sources are generally reionized earlier than large-scale regions far from sources. However, the AM-M and AM-H images appear too grainy. The low-density regions just outside of collapsed regions are assigned low redshifts and considered to be reionized late despite their close proximity to the radiation sources. We can possibly improve these results by smoothing the matter density and halo density fields prior to abundance matching.

Figure \ref{fig:zre_cc} shows the reionization-redshift autopower spectra $\Delta_\mathrm{zz}^2(k)$ and cross correlations relative to the RadHydro normalized redshift $\delta_\mathrm{z}$ (Eq.~\ref{eqn:deltaz}). We only show Sims 0 and 2 for clarity. The AM-R power spectra are most similar to the simulation results, rising in power until a characteristic peak scale and then declining in amplitude with $k$. However, the AM-M and AM-H power spectra generally always increase with $k$, similar to the mass density (Fig.~\ref{fig:rho_cc}) and halo density power spectra (Fig.~\ref{fig:halo_cc}). The AM-R models are the least biased with $b_\mathrm{zz} \approx 1$ for $k \lesssim 1\ h/\mathrm{Mpc}$, while the AM-M and AM-H biases significantly deviate from unity. Furthermore, the AM-R models have improved cross correlation $r_\mathrm{zz}$ that approach unity on large scales. The bias and stochasticity on small scales can be attributed to three main factors. In the RadHydro simulations, the episodic star formation and spatially varying mean free path are not accounted for. Furthermore, the simulations and models have differences in smoothing near the grid scale.

In AMBER, we choose to abundance match using the radiation field evaluated at the redshift midpoint. In future work, we can improve the spatial accuracy by incorporating varying mean free paths \citep[e.g.][]{2021ApJ...917L..37C, 2021arXiv210309821D}. We can also perform the abundance matching tomographically using multiple redshifts.
\clearpage
\section{Methods Comparison} \label{sec:comparison}

We compare AMBER against two other semi-numerical methods using RadHydro Sim 1 as reference. We continue to use the fiducial resolution with $64^3$ mesh with cell size $l_\mathrm{c} = 0.8\ h^{-1}\mathrm{Mpc}$, which is comparable to the typical resolution for semi-numerical methods. In this section, we first summarize the methods (Sec.~\ref{sec:21cmfast_compare} and \ref{sec:rls_compare}), and then compare their reionization histories (Sec.~\ref{sec:xi_compare}) and reionization-redshift fields (Sec.~\ref{sec:zre_compare}).

\subsection{21cmFAST} \label{sec:21cmfast_compare}

The 21cmFAST code \citep{2011MNRAS.411..955M}, like the majority of semi-numerical methods, is based on the ESF for reionization \citep[e.g.][]{2004ApJ...613....1F}. It generates the density, ionization, velocity, and spin temperature fields to compute the 21cm brightness temperature given the astrophysical parameters that control the ionizing, Ly$\alpha$, and X-ray backgrounds. This method has been used widely to model and study the EoR through the 21cm signal \citep[e.g][]{2017MNRAS.472.2651G} and CMB \citep[e.g.][]{2012MNRAS.422.1403M}.

We use the original version with three free parameters:~ionizing efficiency $\zeta$, minimum halo mass $M_\mathrm{min}$, and maximum filter scale $R_\mathrm{max}$, because this parameterization is common amongst ESF codes and has been extensively tested and applied. \citet{2019MNRAS.484..933P} have developed a new version with eight free parameters, allowing additional parameterization of high-redshift galaxy properties. \citet{2021arXiv210309821D} have proposed improving $R_\mathrm{max}$ with two physically motivated prescriptions of the mean free path of ionizing photons. To our knowledge, the newer versions have not yet been tested against RT simulations, and searching the high-dimensional parameter space is beyond the scope of our basic comparison.

By default, 21cmFAST uses 64 times as many 1LPT particles as mesh cells for constructing density and velocity fields. For this comparison, $256^3$ particles are assigned to a $64^3$ mesh using the NGP scheme, but there is significant smoothing near the grid scale because deconvolution is not performed. To achieve the same effective resolution, 21cmFAST would require at least 8 times as many mesh cells, 512 times as many particles, and over 500 times as much memory as AMBER. It also uses the alternative ESF-E version for the collapsed fraction (Sec.~\ref{sec:halo_amber}), but it does not explicitly compute the halo mass density field.

\subsection{RLS} \label{sec:rls_compare}

The Reionization on Large Scales \citep[RLS;][]{2013ApJ...776...81B} method is not based on the ESF, but is a parametric approach motivated by and calibrated with previous RadHydro simulations. It provides a parametric approach to map higher-resolution, smaller-volume radiation-hydrodynamic simulations onto lower-resolution, larger-volume N-body simulations. This method has been applied to model and study the patchy Thomson optical depth \citep{2013ApJ...776...82N}, patchy KSZ effect \citep{2013ApJ...776...83B}, and the 21cm lightcone effect \citep{2014ApJ...789...31L}.

The reionization-redshift field is found to be highly correlated with the matter density field down to Mpc scales and related using a first-order, scale-dependent bias. In Fourier space, the matter overdensity field $\delta(\boldsymbol{k})$ at the mean redshift $\bar{z}$ is first smoothed with a spherical tophat filter $W_\mathrm{th}(k)$ with radius $R_\mathrm{th} = 1\ h^{-1}\mathrm{Mpc}$ and then multiplied by a bias function $b_\mathrm{zm}(k)$ to produce the normalized redshift field,
\begin{equation} \label{eqn:deltazrls}
    \delta_\mathrm{z}(\boldsymbol{k}) = b_\mathrm{zm}(k)\delta(\boldsymbol{k})W_\mathrm{th}(k) .
\end{equation}
The bias function is parameterized as
\begin{equation} \label{eqn:bmzrls}
    b_\mathrm{zm}(k) = \frac{b_\mathrm{RLS}}{(1 + k/k_\mathrm{RLS})^{a_\mathrm{RLS}}} ,
\end{equation}
where the normalization is fixed at $b_\mathrm{RLS} = 1/\delta_\mathrm{c} = 0.593$ based on analytical models \citep{2004ApJ...609..474B}, while the scale $k_\mathrm{RLS}$ and slope $a_\mathrm{RLS}$ are allowed to vary. Finally, the reionization-redshift field is given by
\begin{equation}
    z_\mathrm{re}(\boldsymbol{x}) = \bar{z} + (1+\bar{z})\delta_\mathrm{z}(\boldsymbol{x}) ,
\end{equation}
where the mean redshift $\bar{z}$ is the volume-averaged value of the reionization-redshift field, and it is not exactly equal to the redshift midpoint $z_\mathrm{mid}$ of either the mass-weighted or volume-weighted ionization fractions. While the RLS parameters can be varied to change the reionization history and process, there is no direct connection between them and the redshift midpoint, duration, asymmetry, minimum halo mass, and radiation mean free path.

\subsection{Reionization History} \label{sec:xi_compare}

\begin{figure*}[t]
\includegraphics[width=0.5\hsize]{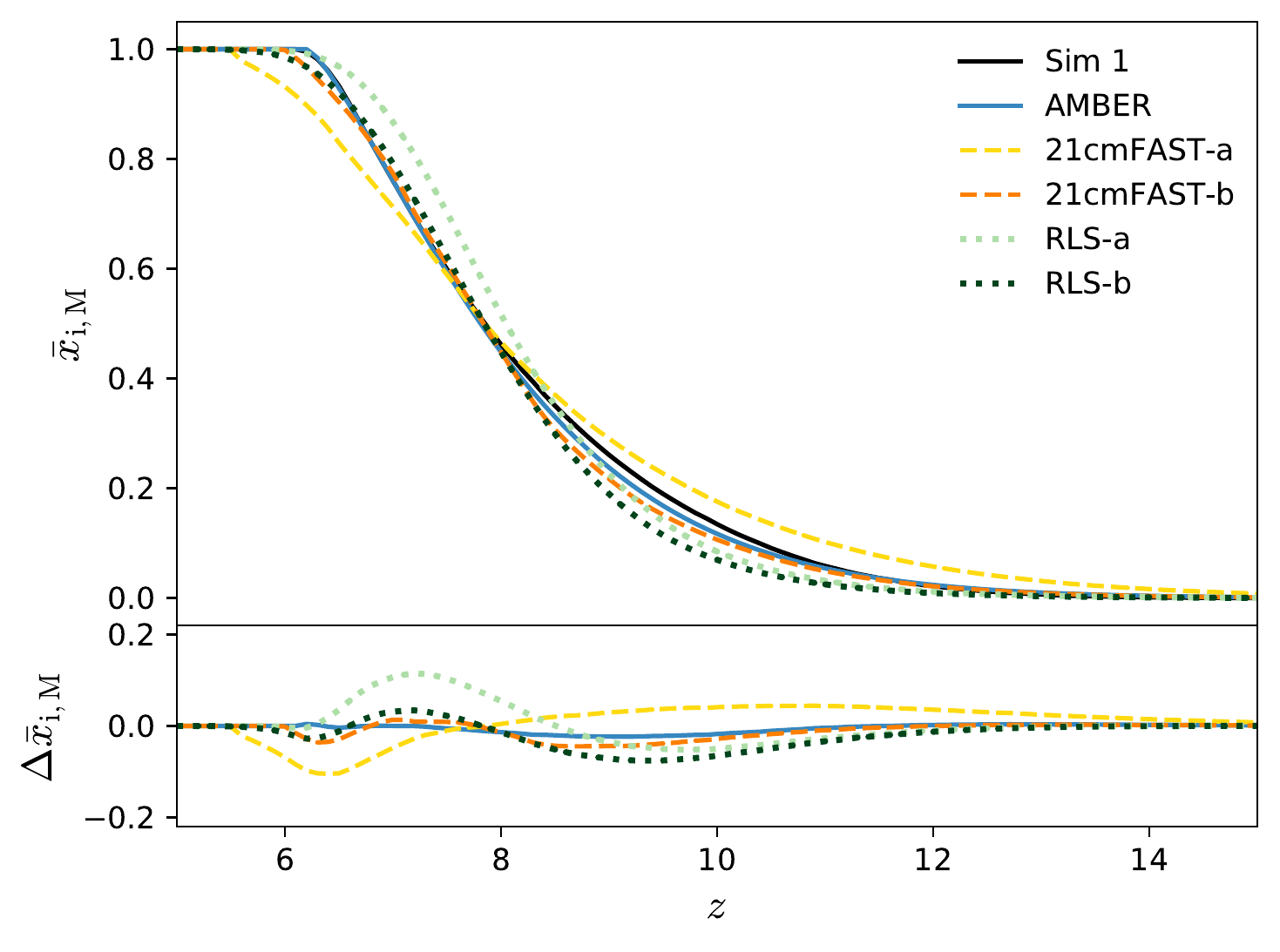}\includegraphics[width=0.5\hsize]{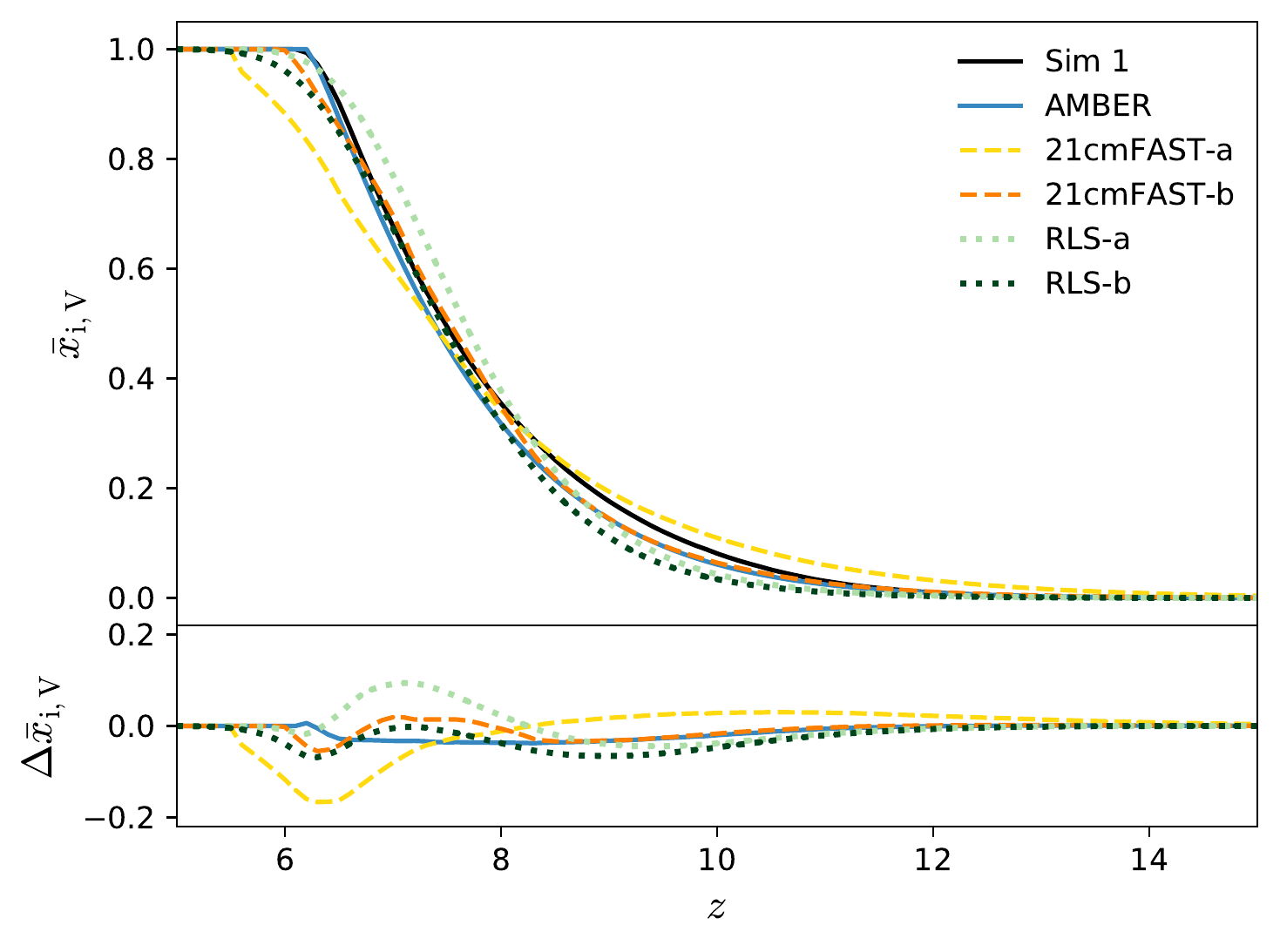}
\caption{{\bf Left:} the evolution of the mass-weighted ionization fractions $\bar{x}_\mathrm{i,M}(z)$ and differences $\Delta\bar{x}_\mathrm{i,M}$ for AMBER, 21cmFAST, and RLS models compared to RadHydro Sim 1. AMBER has almost exact agreement, 21cmFAST-b and RLS-b have similarly good agreement, while 21cmFAST-a and RLS-a have the worst agreement. {\bf Right:} the volume-weighted ionization fractions $\bar{x}_\mathrm{i,V}(z)$ are lower than $\bar{x}_\mathrm{i,M}(z)$ at any given redshift, as higher-density regions near sources are generally reionized earlier than lower-density regions on large scales.}
\label{fig:xi_compare}
\end{figure*}

Figure \ref{fig:xi_compare} compares the mass-weighted ionization fractions $\bar{x}_\mathrm{i,M}(z)$ and volume-weighted ionization fractions $\bar{x}_\mathrm{i,V}(z)$ from the models with the simulation. For AMBER, we directly input the redshift midpoint $z_\mathrm{mid}=7.85$, duration $\Delta_\mathrm{z}=4.73$, and asymmetry $A_\mathrm{z}=2.35$ parameter values. These are calculated for RadHydro Sim 1 at the fiducial resolution using Equations \ref{eqn:duration} and \ref{eqn:asymmetry} on $\bar{x}_\mathrm{i,M}(z)$. While AMBER uses the mass-weighted parameters by construction, it almost exactly reproduces both the mass-weighted ionization fraction with $|\Delta\bar{x}_\mathrm{i,M}|_\mathrm{max} = 0.01$, and volume-weighted ionization fraction with $|\Delta\bar{x}_\mathrm{i,V}|_\mathrm{max} = 0.03$.

There are two models for 21cmFAST. In the first model (21cmFAST-a), we vary the ionization efficiency and find the best-fit $\zeta = 7.5$ for matching $z_\mathrm{mid}$. The minimum halo mass is kept fixed at $M_\mathrm{min} = 10^8\ h^{-1}M_\odot$ to be consistent with the SCORCH galaxy models used in the RadHydro simulations. This model has a more extended reionization history, overestimates the duration by 35\%, underestimates the asymmetry by 6\%, and has $|\Delta\bar{x}_\mathrm{i,M}|_\mathrm{max} = 0.10$ and $|\Delta\bar{x}_\mathrm{i,V}|_\mathrm{max} = 0.17$. In the second model (21cmFAST-b), we find best-fit $\zeta = 22.9$ and $M_\mathrm{min}=1.8\times 10^{9}\ h^{-1}M_\odot$ for matching both $z_\mathrm{mid}$ and $\Delta_\mathrm{z}$. This model underestimates the asymmetry by 14\% and has $|\Delta\bar{x}_\mathrm{i,M}|_\mathrm{max} = 0.04$ and $|\Delta\bar{x}_\mathrm{i,V}|_\mathrm{max} = 0.06$. Note that the best-fit $M_\mathrm{min}$ is now $\approx 20$ times larger and inconsistent with that in the RadHydro simulations. This complicates the interpretation of the inferred minimum halo mass for galaxy formation when comparing observations with the original version of 21cmFAST and other similar ESF models. The newer version \citep{2019MNRAS.484..933P} with additional parameterization for high-redshift galaxy properties will likely provide better agreement, but it requires varying eight free parameters.

There are also two models for the RLS method. The first model (RLS-a) strictly follows \citet{2013ApJ...776...81B}. We fit the bias data from RadHydro Sim 1 and obtain $k_\mathrm{RLS}=2.09\ h/\mathrm{Mpc}$ and $a_\mathrm{RLS}=1.71$ when $b_\mathrm{RLS}$ is set to $b_\mathrm{B13} = 0.593$. In addition, we choose $\bar{z}=7.84$ to be the same value from the simulation. The fitted function underestimates the large-scale bias, and the best-fit $k_\mathrm{RLS}$ and $a_\mathrm{RLS}$ are significantly different from the fiducial values of $k_\mathrm{B13}=0.185\ h/\mathrm{Mpc}$ and $a_\mathrm{B13} = 0.564$ in \citet{2013ApJ...776...81B}. This model increases $z_\mathrm{mid}$ by 0.19, underestimates $\Delta_\mathrm{z}$ by 17\%, underestimates $A_\mathrm{z}$ by 25\%, and has $|\Delta x_\mathrm{i,M}|_\mathrm{max} = 0.11$ and $|\Delta x_\mathrm{i,V}|_\mathrm{max} = 0.09$. In the second model (RLS-b), we fit for all three bias parameters and obtain $b_\mathrm{RLS}=0.895$, $k_\mathrm{RLS}=0.325\ h/\mathrm{Mpc}$, and $a_\mathrm{RLS}=0.816$. While $k_\mathrm{RLS}$ and $a_\mathrm{RLS}$ are now more similar to $k_\mathrm{B13}$ and $a_\mathrm{B13}$, the normalization $b_\mathrm{RLS}$ is 51\% larger than the previously adopted $b_\mathrm{B13}$. We also choose $\bar{z}=7.61$ in order to match $z_\mathrm{mid}$, but note the difference of 0.24 between them. This model underestimates $\Delta_\mathrm{z}$ by 16\%, underestimates $A_\mathrm{z}$ by 30\%, and has $|\Delta x_\mathrm{i,M}|_\mathrm{max} = 0.08$ and $|\Delta x_\mathrm{i,V}|_\mathrm{max} = 0.07$. Note that the RLS method tends to produce more symmetric reionization histories, and it is difficult to produce larger $A_\mathrm{z}$ values just by varying the free parameters.

\subsection{Reionization-redshift Field} \label{sec:zre_compare}

\begin{figure*}[t]
\includegraphics[width=\textwidth]{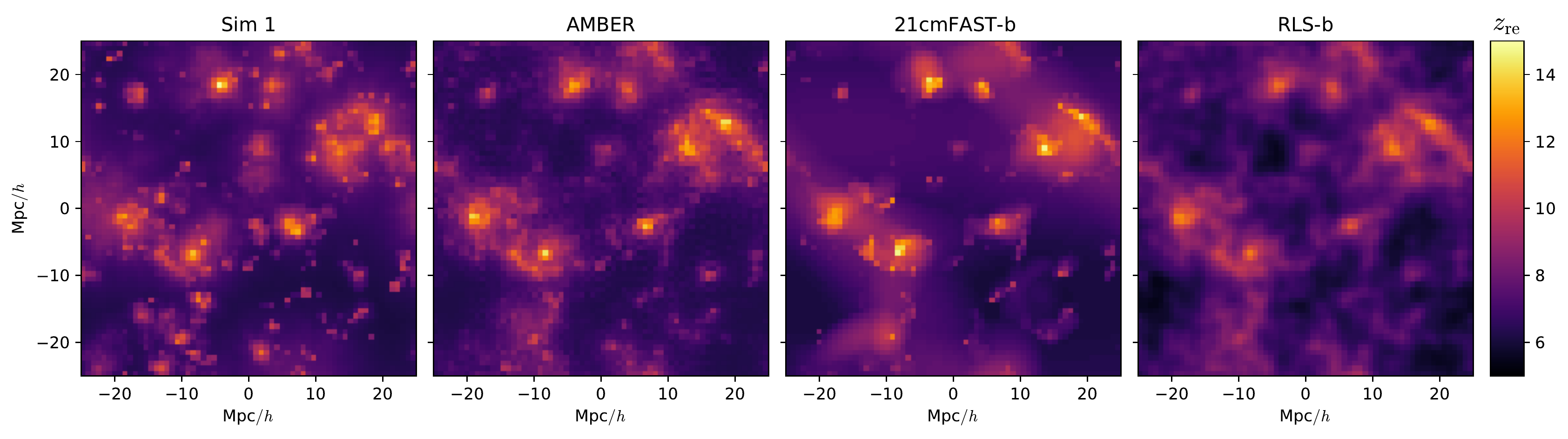}
\caption{Visualization of the reionization-redshift fields $z_\mathrm{re}(\boldsymbol{x})$ from RadHydro Sim 1, AMBER, 21cmFAST, and RLS. Each image is $64 \times 64$ pixels from a slice that is $50\, h^{-1}\mathrm{Mpc} \times 50\, h^{-1}\mathrm{Mpc}$ with a thickness of $\sim 1\, h^{-1}\mathrm{Mpc}$. The AMBER image most closely resembles RadHydro. The 21cmFAST-b image has more stochasticity, while RLS-b has more smoothing.}
\label{fig:zreimg_compare}
\end{figure*}

\begin{figure}[t]
\includegraphics[width=\linewidth]{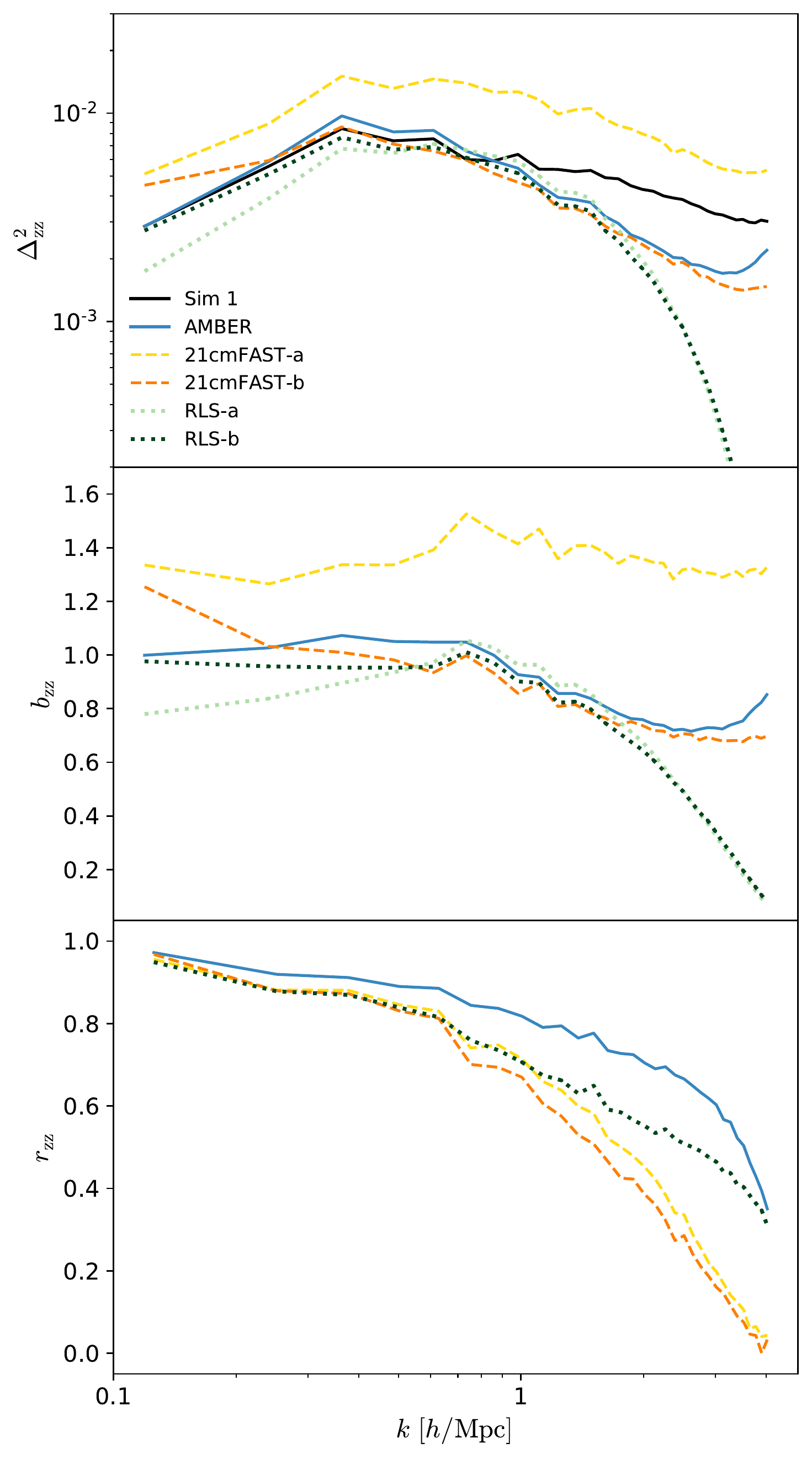}
\caption{{\bf Top:} autopower spectra of the reionization-redshift fields from RadHydro Sim 1, AMBER, 21cmFAST, and RLS. All of the spectra peak at a characteristic scale that depends on the redshift midpoint $z_\mathrm{mid}$. {\bf Center:} the redshift bias $b_\mathrm{zz}$ between the models and simulation show that AMBER and RLS-b are the least biased. {\bf Bottom:} the cross correlations $r_\mathrm{zz}$ show that AMBER is the most correlated, followed by RLS, while 21cmFAST is the most stochastic.}
\label{fig:zrecc_compare}
\end{figure}

Figure \ref{fig:zreimg_compare} is a visualization of the reionization-redshift fields $z_\mathrm{re}(\boldsymbol{x})$ and Figure \ref{fig:zrecc_compare} shows the reionization-redshift autopower spectra $\Delta_\mathrm{zz}^2(k)$ and cross correlations relative to RadHydro Sim 1. For AMBER, we vary the radiation mean free path and obtain $l_\mathrm{mfp} = 3.2\ h^{-1}\mathrm{Mpc}$ when matching the auto-power spectrum on the largest scales. This produces a reionization-redshift field that is visually and statistically in strongest agreement out of all the models. The redshift bias $b_\mathrm{zz}(k)$ and cross correlation $r_\mathrm{zz}(k)$ show that AMBER is both the least biased and most correlated with the RadHydro simulation.

For 21cmFAST, we can also vary the ESF maximum filter scale $R_\mathrm{max}$. It is recommended that $R_\mathrm{max} \ge 20\ h^{-1}\mathrm{Mpc}$, but we are limited to $R_\mathrm{max} \leq 25\ h^{-1}\mathrm{Mpc}$. The radius of the smoothing sphere cannot be larger than half of the periodic box length otherwise it would wrap around itself. We find no improvement over this narrow range of filtering scales. 21cmFAST-b does better than 21cmFAST-a because we are able to match both $z_\mathrm{mid}$ and $\Delta_\mathrm{z}$. Even then it differs more from RadHydro than either AMBER or RLS. The reionization-redshift image shows noticeable differences near biased sources, and this coincides with the redshift bias $b_\mathrm{zz}(k)$ being larger than unity by $>0.2$ on the largest scale. The redshift cross correlation $r_\mathrm{zz}(k)$ drops to 0.5 by $k \sim 1\ h/\mathrm{Mpc}$ and to zero at the grid Nyquist frequency. The stochasticity may be due to using the NGP scheme for the matter density field, the ESF-E version for the collapsed fraction, and the semi-numerical approach for the ionization fraction field. Again, the newer version of 21cmFAST will likely provide better agreement, but it requires searching an eight-dimensional parameter space.

The RLS method is a parametric approach to constructing the reionization-redshift field, so it is not surprising to see that it is visually and statistically in good agreement. On large scales, RSL-b does better than RLS-a because the bias normalization is allowed to vary beyond the fixed value of $b_\mathrm{B13} = 0.593$. On small scales, both models have significant smoothing because of the spherical tophat filter applied to the overdensity field. In \citet{2013ApJ...776...81B}, this implementation choice was made because the linear, deterministic biasing relation, $\delta_\mathrm{z}(\boldsymbol{k}) \propto \delta(\boldsymbol{k})$, breaks down on scales $k \gtrsim 1\ h/\mathrm{Mpc}$. RLS has more stochasticity than AMBER, but less than 21cmFAST. The gradual deviation of the redshift cross correlation $r_\mathrm{zz}(k)$ from unity reflects the decreasing correlation between the reionization-redshift and matter density fields towards smaller scales.

More studies and comparisons are required to understand the advantages, disadvantages, and appropriate utilization of the various semi-numerical methods. We emphasize that a variety of independent approaches are crucial for theoretical and inference studies of the EoR.

\section{Results} \label{sec:results}

We now present some basic results from AMBER and leave further applications for upcoming work. For example, \citet{2022arXiv220304337C} models and studies the patchy KSZ effect, which is a promising probe of the EoR. In this section, we study the parameter dependence of the reionization-redshift field (Sec.~\ref{sec:zre_results}) by varying the redshift midpoint, duration, asymmetry, minimum halo mass, and radiation mean free path.

We run AMBER models, each with $2048^3$ particles and $2048^3$ cells in a comoving box of side length $L = 2\ h^{-1}\mathrm{Gpc}$. For the fiducial model, we use $z_\mathrm{mid} = 8.0$, $\Delta_\mathrm{z} = 4.0$, $A_\mathrm{z} = 3.0$, $M_\mathrm{min} = 10^8\ h^{-1}M_\odot$, and $l_\mathrm{mfp} = 3.0\ h^{-1}\mathrm{Mpc}$. In addition, we include a lower and higher value when we vary each parameter one at a time. Each simulation takes approximately 20 CPU hours when run in parallel with 32 to 64 cores. We discuss the code scaling performance in Appendix \ref{app:scaling}.

\subsection{Reionization-redshift Fields} \label{sec:zre_results}

\begin{figure*}[t]
\includegraphics[width=\textwidth]{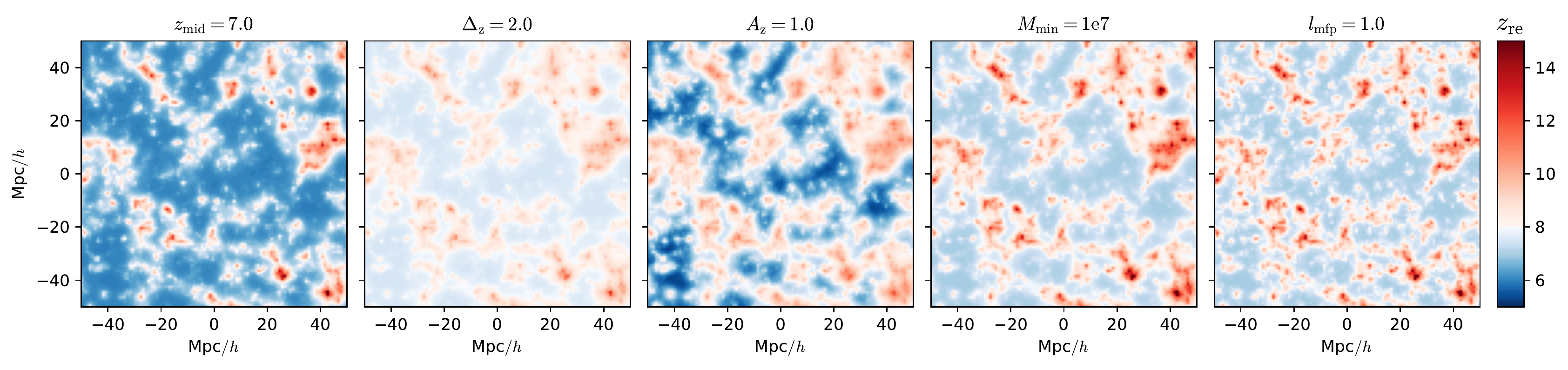}
\includegraphics[width=\textwidth]{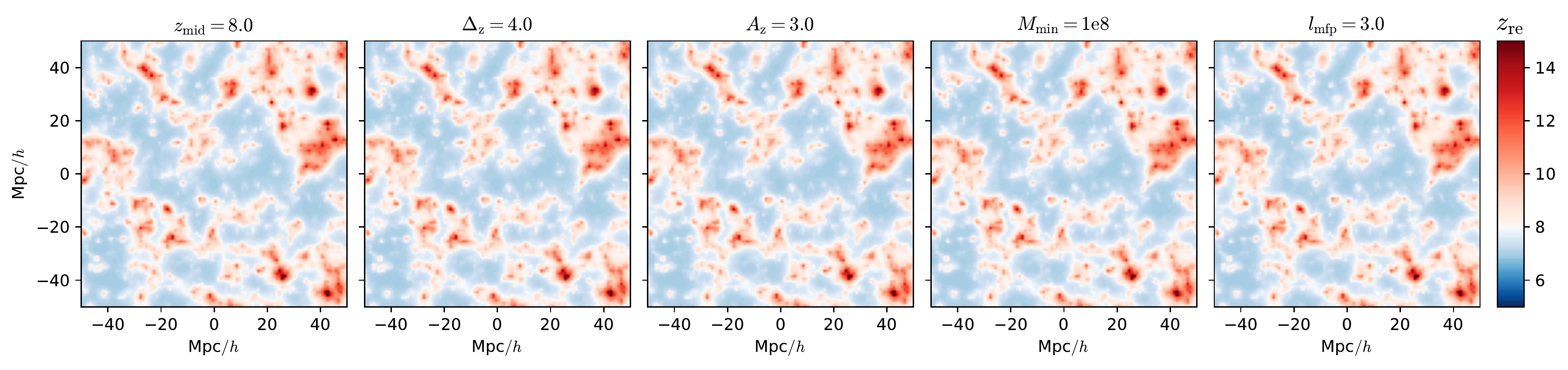}
\includegraphics[width=\textwidth]{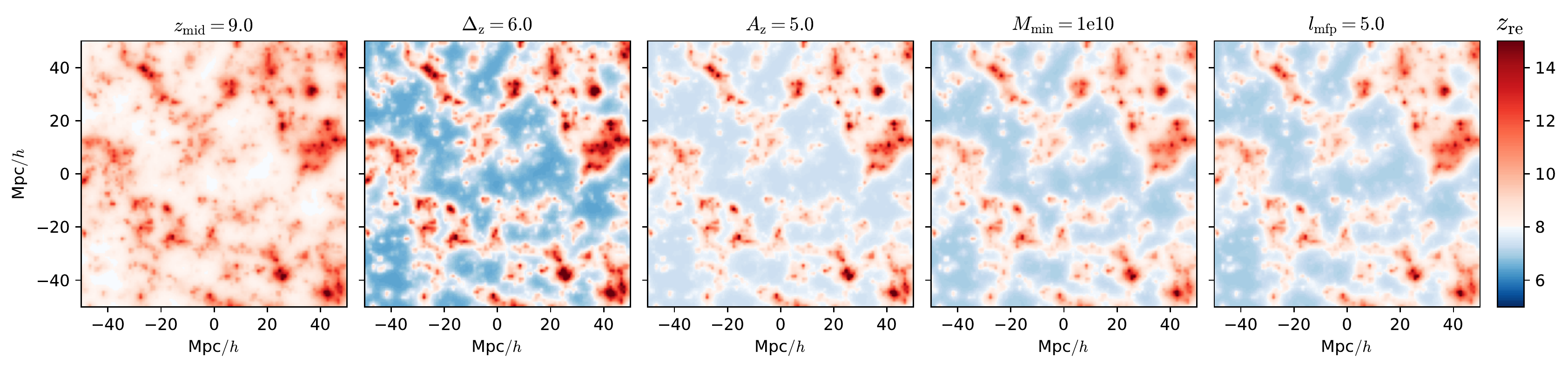}
\caption{Visualization of the reionization-redshift fields $z_\mathrm{re}(\boldsymbol{x})$ when the redshift midpoint $z_\mathrm{mid}$, duration $\Delta_\mathrm{z}$, asymmetry $A_\mathrm{z}$, minimum halo mass $M_\mathrm{min}$, and radiation mean free path $l_\mathrm{mfp}$ are varied. Each column shows when a given parameter is varied, while the middle row shows to the same fiducial model. The divergent color scheme centered at the fiducial $z_\mathrm{mid} = 8.0$ is used to distinguish between regions that are ionized earlier and later. Each image is $100 \times 100$ pixels from a subsection that is $100\, h^{-1}\mathrm{Mpc} \times 100\, h^{-1}\mathrm{Mpc}$ with a thickness of $\sim 1\, h^{-1}\mathrm{Mpc}$.}
\label{fig:zreimg_results}
\end{figure*}

\begin{figure*}[t]
\includegraphics[width=\linewidth]{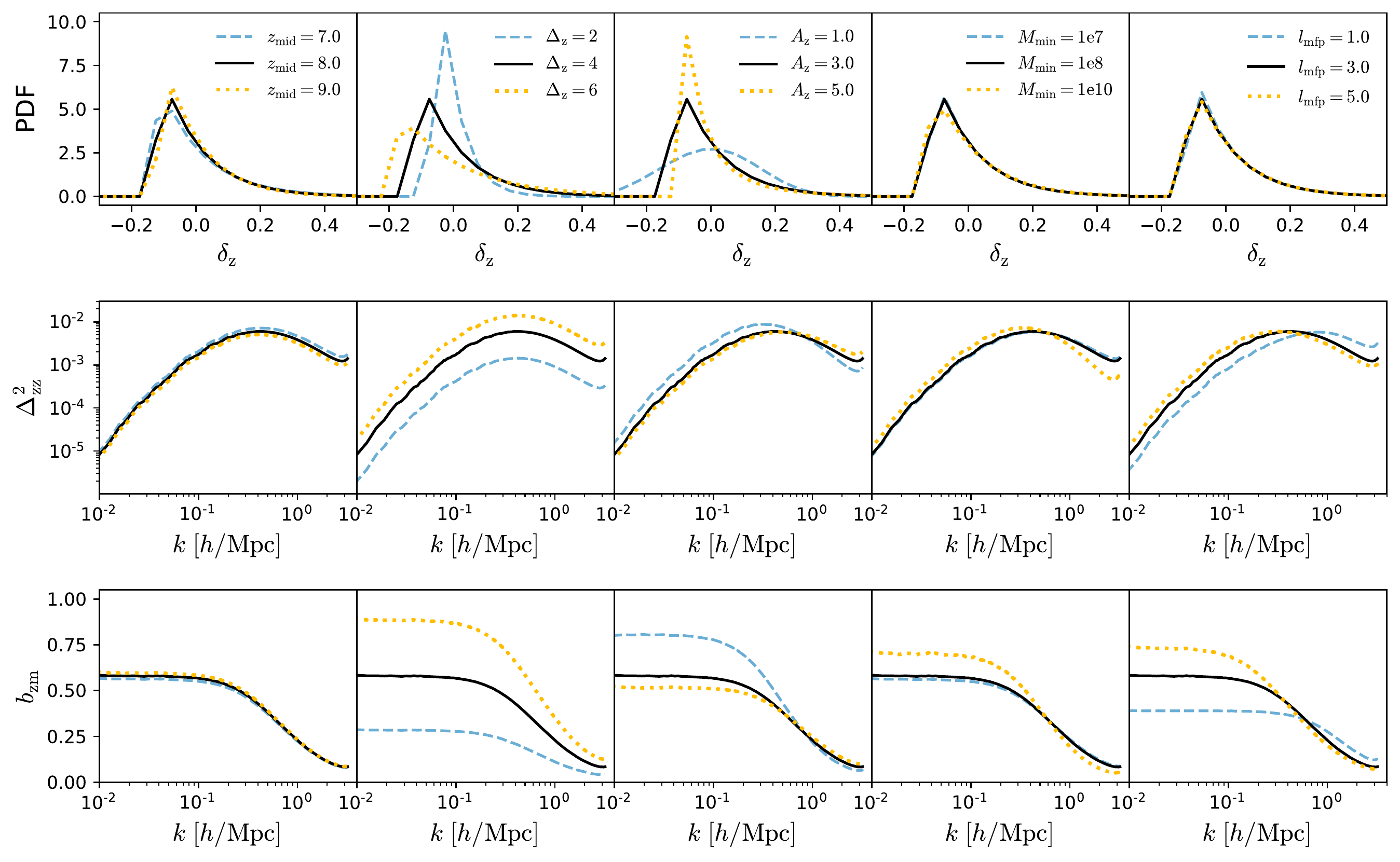}
\caption{{\bf Top:} probability distribution functions of the normalized redshift $\delta_\mathrm{z}$ when the redshift midpoint $z_\mathrm{mid}$, duration $\Delta_\mathrm{z}$, asymmetry $A_\mathrm{z}$, minimum halo mass $M_\mathrm{min}$, and radiation mean free path $l_\mathrm{mfp}$ are varied. {\bf Center:} autopower spectra of the reionization-redshift fields. {\bf Bottom:} the redshift-matter bias $b_\mathrm{zm}$ approaches a constant value on large scales and decreases in value on small scales.}
\label{fig:zre_results}
\end{figure*}

Figure \ref{fig:zreimg_results} is a visualization of the reionization-redshift fields. Each image is $100 \times 100$ pixels from a subsection that is $100\, h^{-1}\mathrm{Mpc} \times 100\, h^{-1}\mathrm{Mpc}$ with a thickness of $\sim 1\, h^{-1}\mathrm{Mpc}$. We use a divergent color scheme centered at the fiducial redshift midpoint $z_\mathrm{mid} = 8.0$ to distinguish between regions that are ionized earlier and later. In general, the higher-density regions near radiation sources are reionized earlier than the lower-density regions farther away on large scales. As a result, a larger fraction of the volume is seen with $z_\mathrm{re} < z_\mathrm{mid}$ and the redshift midpoint of the volume-weighted ionization fraction comes later.

Figure \ref{fig:zre_results} shows the one-point PDF of the normalized redshift $\delta_\mathrm{z}$ (Eq.~\ref{eqn:deltaz}), reionization-redshift auto-power spectra,  and cross correlations relative to the matter density field at the redshift midpoint. In general, the PDF$(\delta_\mathrm{z})$ are skewed Gaussians except for the symmetric case with $A_\mathrm{z}=1.0$. The reionization-redshift power spectrum $\Delta_\mathrm{zz}(k)$ first rises in power until a characteristic peak scale and then declines in amplitude with $k$. The redshift-matter bias $b_\mathrm{zm}(k)$ asymptotically approaches a constant value towards large scales and declines monotonically with $k$.

\subsubsection{Redshift Midpoint} \label{sec:zmid_results}

As the redshift midpoint decreases over the range $7.0 \leq z_\mathrm{mid} \leq 9.0$, we see overall later reionization in Figure \ref{fig:zreimg_results}. The one-point PDF($\delta_\mathrm{z}$) becomes wider, the two-point power spectrum $\Delta_\mathrm{zz}^2(k)$ increases overall in amplitude, but the redshift-matter bias $b_\mathrm{zm}(k)$ remains almost constant in Figure \ref{fig:zre_results}. In Equation \ref{eqn:deltaz} for the normalized redshift $\delta_\mathrm{z}$, the numerator ($z_\mathrm{re} - \bar{z}_\mathrm{re}$) remains approximately constant, but the denominator ($1+\bar{z}_\mathrm{re}$) decreases. Therefore, $\delta_\mathrm{z}$ grows by a factor of $\bar{a}_\mathrm{re} = 1/(1 + \bar{z}_\mathrm{re})$, whereas the matter overdensity $\delta$ grows by a factor $D(z_\mathrm{mid}) \approx 1/(1 + z_\mathrm{mid})$. We find that the normalized redshift grows similarly to the matter overdensity as the bias is only $\approx5\%$ lower for $z_\mathrm{mid} = 7.0$ than for $z_\mathrm{mid} = 9.0$. Note that while the radiation field also grows like the matter density, the abundance matching is insensitive to the overall amplitude.

\subsubsection{Duration} \label{sec:Dz_results}

As the duration increases over the range $2.0 \leq \Delta_\mathrm{z} \leq 6.0$, we see a larger range of values in the reionization-redshift images. The PDF($\delta_\mathrm{z}$) has larger variance and the power spectrum increases overall in amplitude. Note that the variance is given by $\sigma_\mathrm{z}^2 = \langle|\delta_\mathrm{z}(\boldsymbol{x})|^2 \rangle = \int\Delta_\mathrm{zz}^2(k)d\ln k$. Therefore, we expect the large-scale bias to scale with $\sigma_\mathrm{z}$, which in turn scales with $\Delta_\mathrm{z}$. We find that the large-scale bias changes by a factor $\approx 3.1$ when the duration increases from $\Delta_\mathrm{z} = 2.0$ to $\Delta_\mathrm{z} = 6.0$.

The large-scale bias can differ significantly from the single value of $b_\mathrm{B13} = 1/\delta_\mathrm{c} = 0.593$ adopted in the RLS method \citep{2013ApJ...776...81B}. \citet{2004ApJ...609..474B} derived this value assuming that fluctuations in the reionization-redshift field are solely determined by fluctuations in the collapsed mass density field from EPS theory. Relating the growth of ionized regions to the growth of dark matter halos would correspond to a particular duration value. However, the reionization history can depend on other nonconstant factors and the large-scale bias can have a range of values in general.

\subsubsection{Asymmetry} \label{sec:Az_results}

As the asymmetry increases over the range $1.0 \leq A_\mathrm{z} \leq 5.0$, the images show a smaller range of redshift values for regions reionized after the midpoint and a larger range of redshift values for the regions reionized earlier than the midpoint. The PDF($\delta_\mathrm{z}$) becomes skewed and the power spectrum changes shape. There is less power on large scales and more power on small scales as the characteristic peak shifts to smaller scales. For a larger $A_\mathrm{z}$ at fixed duration, the first portion ($z_\mathrm{mid} < z < z_\mathrm{ear}$) of the reionization history becomes longer, while the second portion ($z_\mathrm{lat} < z < z_\mathrm{mid}$) becomes shorter. We have just seen that increasing (decreasing) the duration increases (decreases) the overall power. We find that lengthening the first half portion slightly increases the power on small scales while shortening the second portion more significantly decreases the power on large scales. Correspondingly, the large-scale bias decreases with more asymmetry in the reionization history.

\subsubsection{Minimum Halo Mass} \label{sec:Mmin_results}

Over the minimum halo mass range $10^7 \leq M_\mathrm{min}/[h^{-1}M_\odot] \leq 10^{10}$, there are only minor changes to the reionization-redshift field. The PDF($\delta_\mathrm{z}$) only has minor changes since the redshift midpoint, duration, and asymmetry are kept fixed. There is minor suppression in $\Delta_\mathrm{zz}^2(k)$ and $b_\mathrm{zm}(k)$ on small scales. For larger $M_\mathrm{min}$, the collapsed fraction decreases and the halo bias increases. Correspondingly, this decreases the amplitude and increases the bias of the radiation intensity field (Eq.~\ref{eqn:intensity}). However, the abundance matching technique is insensitive to the overall normalization of $J_\nu$ and to the overall normalization of the fluctuating component $\delta_\mathrm{J}$. The abundance matching only depends on the relative ranking of the field values.

On one hand, varying the minimum halo mass and halo density field will generally change the reionization history and process if astrophysical parameters such as the star formation efficiency, photon production efficiency, and radiation escape fraction are kept fixed. On the other hand, varying these astrophysical terms, which can be functions of mass and redshift, can offset the changes from varying $M_\mathrm{min}$ and produce the same reionization history. With AMBER, we find that when the redshift midpoint, duration, asymmetry, and radiation mean free path are kept fixed, then the reionization-redshift field has only minor dependence on the minimum halo mass. For certain applications, it may then be possible to ignore $M_\mathrm{min}$ and reduce the number of free parameters.

\subsubsection{Radiation Mean Free Path} \label{sec:lmfp_results}

As the radiation mean free path increases over the range $1.0 \leq l_\mathrm{mfp}/[h^{-1}\mathrm{Mpc}] \leq 5.0$, we see more spatial coherence in the reionization-redshift images. For a larger $l_\mathrm{mfp}$, there is a larger coherence length and less small-scale structure in the radiation intensity field used for abundance matching. The PDF($\delta_\mathrm{z}$) only has minor changes since the redshift midpoint, duration, and asymmetry are kept fixed. The reionization-redshift power spectrum changes shape though, having more power on large scales and less on small scales. Increasing $l_\mathrm{mfp}$ produces relatively larger ionized regions, resulting in the characteristic peak shifting to larger scales.

In the RLS method, \citet{2013ApJ...776...81B} change the duration by varying $a_\mathrm{RLS}$ and $k_\mathrm{RLS}$. But without varying $b_\mathrm{RLS}$, they also change the correlation length of ionized regions at the same time. More specifically, they obtain smaller correlation lengths for longer durations. With AMBER, we find longer $\Delta_\mathrm{z}$ increases the bias, while shorter $l_\mathrm{mfp}$ decreases the bias. This explains the inverse relationship between the duration and correlation length when $b_\mathrm{RLS}$ is fixed in the RLS method. AMBER has the flexibility to change $\Delta_\mathrm{z}$ and $l_\mathrm{mfp}$ independently.
\section{Conclusion} \label{sec:conclusion}

AMBER is a semi-numerical code for modeling the cosmic dawn and EoR. The new algorithm is not based on the ESF for directly predicting the ionization fraction field, but takes the novel approach of calculating the reionization-redshift field $z_\mathrm{re}(\boldsymbol{x})$ assuming that the hydrogen gas encountering higher radiation intensity is photoionized earlier. Redshift values are assigned while matching the abundance of ionized mass according to a given mass-weighted ionization fraction $x_\mathrm{i}(z)$. The code has the unique advantage of allowing users to directly specify the reionization history through the redshift midpoint $z_\mathrm{mid}$, duration $\Delta_\mathrm{z}$, and asymmetry $A_\mathrm{z}$ input parameters \citep{2018ApJ...858L..11T}. The reionization process is further controlled through the minimum halo mass $M_\mathrm{min}$ for galaxy formation and the radiation mean free path $l_\mathrm{mfp}$ for RT.

We implement improved methods for modeling the large-scale structure and constructing density, velocity, halo, and radiation fields, which are essential components for modeling reionization observables. 2LPT (vs 1LPT) is used to efficiently evolve the matter distribution in the moderately nonlinear regime. The TSC (vs NGP or CIC) PM assignment scheme with interlacing and deconvolution produces density and velocity fields that are in excellent agreement with RadHydro simulations \citep{2019ApJ...870...18D}. ESF-L (vs ESF-E) accurately and robustly produces halo mass density fields compared to N-body simulations \citep{2015ApJ...813...54T}. The radiation intensity field is computed quickly using FFTs to convolve the source field with the RT kernel. The interlacing and deconvolution reduce aliasing and smoothing, and thereby improve power spectrum estimation \citep[e.g.][]{2016MNRAS.460.3624S}. These methods can also be used to improve other semi-numerical and RT algorithms. The AMBER program modules can be adapted and plugged into other codes.

We compare AMBER with two other semi-numerical methods, 21cmFAST \citep{2011MNRAS.411..955M} and the RLS method \citep{2013ApJ...776...81B}, and find that our code more accurately reproduces the results from RadHydro simulations. AMBER has the flexibility to directly match the simulated reionization history $x_\mathrm{i}(z)$, while the other two methods require running multiple models before finding the best fit. It also produces reionization-redshift fields $z_\mathrm{re}(\boldsymbol{x})$ that are the least biased and most correlated with the simulations. The RLS method has more smoothing, while 21cmFAST has more stochasticity. More studies and comparisons are required to understand the advantages, disadvantages, and appropriate utilization of the various semi-numerical methods. We reiterate that a variety of independent approaches are crucial for theoretical and inference studies of the EoR.

We study the parameter dependence of the reionization-redshift field by varying the redshift midpoint, duration, asymmetry, minimum halo mass, and radiation mean free path. As $z_\mathrm{mid}$ decreases, there is overall later reionization, the PDF($\delta_\mathrm{z}$) becomes wider, the power spectrum $\Delta_\mathrm{zz}^2(k)$ increases overall in amplitude, but the redshift-matter bias $b_\mathrm{zm}(k)$ remains approximately constant. As $\Delta_\mathrm{z}$ increases, there is more extended reionization, the redshift distribution becomes wider, and both the power spectrum and bias increase overall in amplitude. As $A_\mathrm{z}$ increases, the reionization history becomes more asymmetric, the redshift distribution is more skewed, and the power spectrum changes shape. Interestingly, the reionization-redshift field is only weakly sensitive to $M_\mathrm{min}$ when the other parameters are fixed. As $l_\mathrm{mfp}$ increases, there is more spatial coherence and less small-scale structure in the radiation intensity and reionization-redshift fields, and the power spectrum changes shape due to the relatively larger ionized regions.

AMBER is initially designed for theoretical and inference studies in which the primary interest is controlling or constraining the reionization history and process. With the reionization-redshift field, the evolution of the neutral hydrogen and ionized electron densities can be readily computed. As a first application, \citet{2022arXiv220304337C} models and studies the imprint of reionization on the CMB through the patchy KSZ effect, which is a promising probe of the EoR. Complementary, the connection between radiation sources and sinks, and the reionization history can be understood using analytical models for the evolution of the mass-weighted ionization fraction \citep[e.g.][]{2020ApJ...905..132C} and radiation-hydrodynamic simulations that solve the complex physics.

In future work, we will include additional physics that will enable more realistic predictions of EoR observables. To improve the source models, we will calibrate the ESF halo collapsed fraction against N-body simulations, and include high-redshift galaxies through abundance matching \citep[e.g.][]{2015ApJ...813...54T}. To improve the radiation and reionization-redshift fields, we will incorporate spatially varying radiation mean free paths \citep[e.g.][]{2021ApJ...917L..37C, 2021arXiv210309821D}, and perform the ionization abundance matching tomographically using multiple redshifts. We will compute the Ly$\alpha$ and X-ray radiation fields for the 21cm signal \citep[e.g.][]{2011MNRAS.411..955M, 2017MNRAS.472.4508S} and calculate the thermal history for the Lyman alpha forest \citep[e.g.][]{2016MNRAS.460.1885U, 2019ApJ...874..154D}.

AMBER is currently parallelized to run on multi-core, shared-memory nodes. It is over four orders of magnitude faster than radiation-hydrodynamic simulations and will efficiently enable large-volume models, full-sky mock observations, and parameter-space studies. The code will be made publicly available to facilitate and transform studies of the EoR.

\begin{acknowledgements}
We thank Nick Gnedin for initial discussions that motivated the development of AMBER. We thank Anson D'Aloisio, Matt McQuinn, and the anonymous referee for providing comments on the manuscript. We also thank Nick Battaglia, Sasha Kaurov, Emiliano Sefusatti, and Qirong Zhu for helpful discussions. The computational work was performed on the Bridges-2 system at the Pittsburgh Supercomputing Center (PSC) using allocations ast190014p and phy210042p. We thank PSC staff Ken Hackworth and T.J.~Olesky for computing support. H.T.~acknowledges support from the NSF AI Institute:~Planning:~Physics of the Future, NSF PHY2020295. R.C.~acknowledges support from NSF AST2007390.
\end{acknowledgements}

\bibliography{amber}{}
\bibliographystyle{aasjournal}

\appendix

\section{Weibull Function} \label{app:weibull}

The reionization history $\bar{x}_\mathrm{i}(z)$ can be more accurately fit with a flexible Weibull function (Eq.~\ref{eqn:weibull}) than a symmetric tanh function \citep[e.g.][]{2008PhRvD..78b3002L}. The Weibull coefficients can be determined by first solving a nonlinear equation for $c_\mathrm{w}$ and then substituting its value into algebraic equations for $a_\mathrm{w}$ and $b_\mathrm{w}$. There are several variations of the root-finding equation, for example:
\begin{equation}
    f(c_\mathrm{w}) = \left(\frac{\ln\bar{x}_\mathrm{ear}}{\ln\bar{x}_\mathrm{mid} }\right)^{1/c_\mathrm{w}} - \left(\frac{z_\mathrm{ear} - a}{z_\mathrm{mid} - a}\right) = 0 ,
\end{equation}
where 
\begin{equation}
    a = \frac{z_\mathrm{lat}|\ln x_\mathrm{mid}|^{1/c_\mathrm{w}} - z_\mathrm{mid}|\ln x_\mathrm{lat}|^{1/c_\mathrm{w}}}{|\ln x_\mathrm{mid}|^{1/c_\mathrm{w}} - |\ln x_\mathrm{lat}|^{1/c_\mathrm{w}}} ,
\end{equation}
which can be solved iteratively using Newton's method with finite differences instead of analytical derivatives (i.e.~secant method). The other two coefficients have algebraic solutions, for example:
\begin{align}
    a_\mathrm{w} & = \frac{z_\mathrm{ear}|\ln x_\mathrm{lat}|^{1/c_\mathrm{w}} - z_\mathrm{lat}|\ln x_\mathrm{ear}|^{1/c_\mathrm{w}}}{|\ln x_\mathrm{lat}|^{1/c_\mathrm{w}} - |\ln x_\mathrm{ear}|^{1/c_\mathrm{w}}}, \\[10pt]
    b_\mathrm{w} & = \frac{z_\mathrm{mid} - a_\mathrm{w}}{|\ln x_\mathrm{mid}|^{1/c_\mathrm{w}}} .
\end{align}
Note that the equations above can also be written in terms of the redshift midpoint, duration, and asymmetry parameters by replacing $z_\mathrm{ear}$ and $z_\mathrm{lat}$ with $\Delta_\mathrm{z}$ and $A_\mathrm{z}$ using Equations \ref{eqn:zear} and \ref{eqn:zlat}.

\section{Scaling Performance} \label{app:scaling}

\begin{figure}[t]
\includegraphics[width=\linewidth]{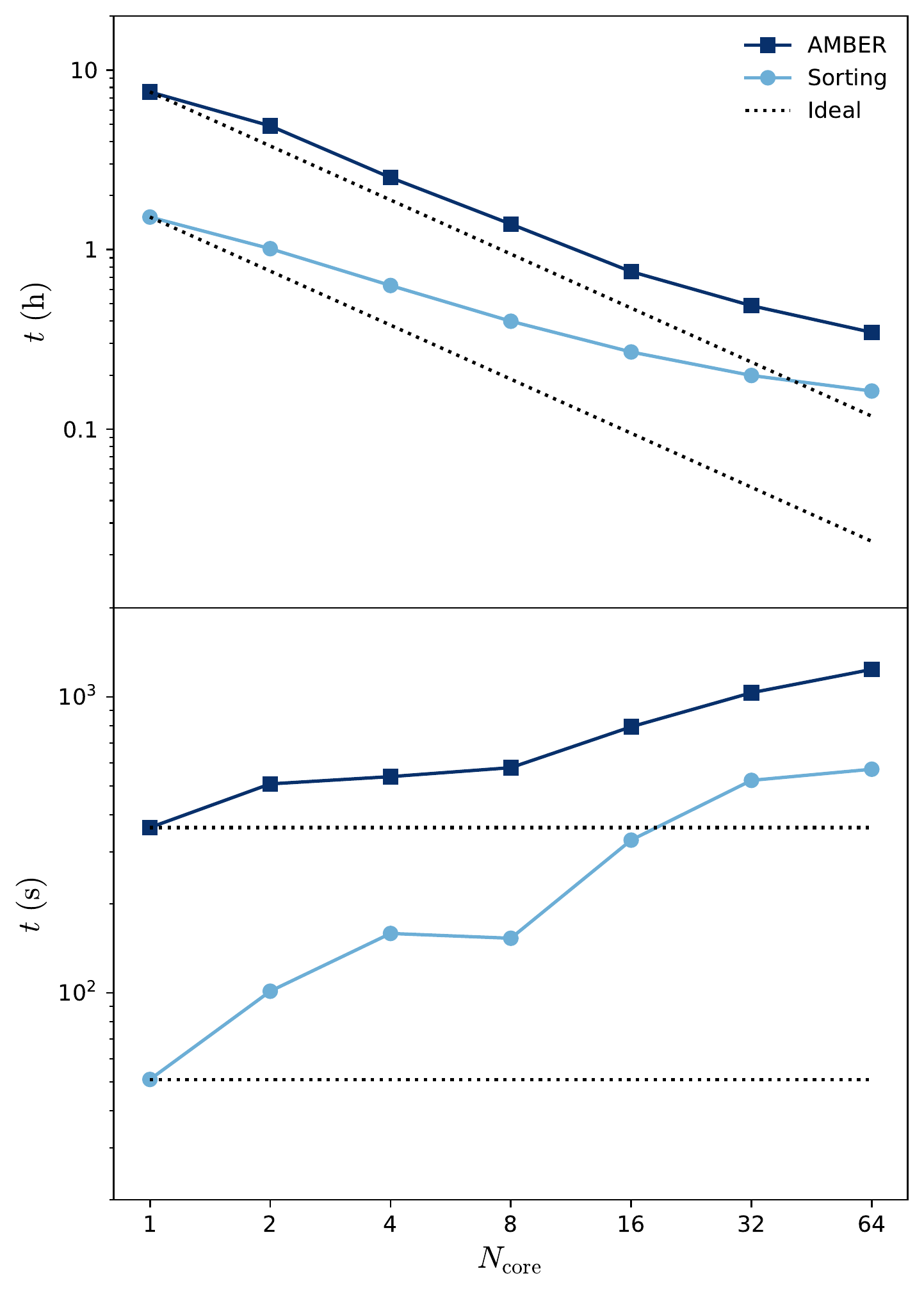}
\caption{{\bf Top:} strong scaling test with a fixed number of $N_\mathrm{part}=2048^3$ particles for each AMBER run. The measured wall time $t$ is compared against the ideal scaling. {\bf Bottom:} weak scaling test with a particle/core ratio $N_\mathrm{part}/N_\mathrm{core} \approx 512^3$.}
\label{fig:scaling}
\end{figure}

AMBER is written in modern Fortran and parallelized using OpenMP to run on multi-core, shared-memory nodes. We perform the following scaling tests on Bridges-2 at the Pittsburgh Supercomputing Center (PSC). An Extreme Memory (EM) node has 4 Intel Xeon Platinum 8260M ``Cascade Lake" CPUs, 24 cores per CPU, 96 cores per node, and a total of 4 TB of memory.

Figure \ref{fig:scaling} shows the results of the scaling tests. For the strong scaling test, AMBER is run with $N_\mathrm{part} = 2048^3$ particles and equal number of mesh cells in a comoving box of side length $L = 2\ h^{-1}\mathrm{Gpc}$. We vary the number of cores $N_\mathrm{core}$ from 1 to 64 and measure the wall time $t(N_\mathrm{core})$.
The speedup is defined as
\begin{equation}
    S = \frac{t(1)}{t(N_\mathrm{core})} ,
\end{equation}
and we obtain $S = 5.4\ (21.8)$ for $N_\mathrm{core} = 8\ (64)$. The deviation from ideal speedup is mostly due to our parallel quicksort, which is not fully optimized yet. The serial sorting takes about $20\%$ of the total wall time, while the parallel sorting takes $29\%\ (47\%) $ for $N_\mathrm{core} = 8\ (64)$. Nonetheless, a typical AMBER model can be quickly run in under an hour wall time with 32 to 64 cores.

For the weak scaling test, we use a particle/core ratio of $N_\mathrm{part}/N_\mathrm{core} \approx 512^3$. We vary $N_\mathrm{core}$ from 1 to 64 and change $N_\mathrm{part}$ accordingly. The efficiency is defined as
\begin{equation}
    E = \frac{t(1)}{t(N_\mathrm{core})} ,
\end{equation}
and we obtain $E = 61\%\ (30\%)$ for $N_\mathrm{core} = 8\ (64)$. Again, the decrease in efficiency is mostly due to our nonoptimal quicksort. The serial sorting takes about $14\%$ of the total wall time, while the parallel sorting takes $27\%\ (47\%) $ for $N_\mathrm{core} = 8\ (64)$. We will improve the parallelization of the algorithm in future work.

\end{document}